\documentclass[%
 aip,
 floatfix,
 amsmath,amssymb,
 reprint,%
]{revtex4-1}

\usepackage{graphicx}
\usepackage{dcolumn}
\usepackage{bm}

\usepackage[utf8]{inputenc}
\usepackage[T1]{fontenc}
\usepackage{mathptmx}
\usepackage{etoolbox}
\usepackage{here}

\makeatletter
\def\@email#1#2{%
 \endgroup
 \patchcmd{\titleblock@produce}
  {\frontmatter@RRAPformat}
  {\frontmatter@RRAPformat{\produce@RRAP{*#1\href{mailto:#2}{#2}}}\frontmatter@RRAPformat}
  {}{}
}%
\makeatother
\begin{document}

\preprint{AIP/123-QED}

\title[Theoretical Characteristics of a three-point Roberts Linkage]
{Theoretical Characteristics of a Three-Point Roberts Linkage}
\author{M. Otsuka}
 \email{munetake.otsuka@grad.nao.ac.jp}
 \affiliation{Graduate School of Science, the University of Tokyo, 7-3-1 Hongō, Bunkyo, Tokyo, 113-0033 Japan}
\affiliation{National Astronomical Observatory of Japan, 2-21-1 Osawa, Mitaka, Tokyo, 181-0015 Japan}
\author{K. Mitsuhashi}
\affiliation{Graduate School of Science, the University of Tokyo, 7-3-1 Hongō, Bunkyo, Tokyo, 113-0033 Japan}
\affiliation{National Astronomical Observatory of Japan, 2-21-1 Osawa, Mitaka, Tokyo, 181-0015 Japan}
\author{R. Takahashi}
\affiliation{National Astronomical Observatory of Japan, 2-21-1 Osawa, Mitaka, Tokyo, 181-0015 Japan}
\affiliation{The Graduate University for Advanced Studies, SOKENDAI, Shonan Village, Hayama, Kanagawa, 240-0193 Japan}
\author{Y. Nishino}
\affiliation{Graduate School of Science, the University of Tokyo, 7-3-1 Hongō, Bunkyo, Tokyo, 113-0033 Japan}
\affiliation{National Astronomical Observatory of Japan, 2-21-1 Osawa, Mitaka, Tokyo, 181-0015 Japan}
\author{Y. Aso}
\affiliation{National Astronomical Observatory of Japan, 2-21-1 Osawa, Mitaka, Tokyo, 181-0015 Japan}
\affiliation{The Graduate University for Advanced Studies, SOKENDAI, Shonan Village, Hayama, Kanagawa, 240-0193 Japan}
\author{T. Tomaru}
\affiliation{Graduate School of Science, the University of Tokyo, 7-3-1 Hongō, Bunkyo, Tokyo, 113-0033 Japan}
\affiliation{National Astronomical Observatory of Japan, 2-21-1 Osawa, Mitaka, Tokyo, 181-0015 Japan}
\affiliation{The Graduate University for Advanced Studies, SOKENDAI, Shonan Village, Hayama, Kanagawa, 240-0193 Japan}
 \homepage{http://www.Second.institution.edu/~Charlie.Author.}
\date{\today}

\begin{abstract}
The Roberts linkage is recognized for enabling long-period pendulum motion in a compact format. Utilizing this characteristic, we are developing a three-point Roberts linkage for vibration isolation systems, with an eye towards its potential contribution to the development of next-generation interferometric gravitational wave antennas. In this article, we derived the equations to determine the essential parameters when using this linkage as a vibration isolation system, namely the equivalent pendulum length and the relationship between translational motion of the center of mass and rigid body rotation, from size parameters. Additionally, we analyzed the behavior in response to various errors.
\end{abstract}

\maketitle

\section{\label{sec:level1}Introduction}

Interferometric gravitational wave detectors track the space-time distortions reflected in the distance changes between mirrors fixed in an inertial  frame, as gravitational waves pass through. Light interference facilitates this precise measurement. Seismic noise, mainly at low frequencies, stands as the major noise source\cite{KAG}, prompting the use of sophisticated vibration isolation systems (VIS) to shield the detectors' test masses from ground vibrations. Enhancing these systems with multiple stages further improves isolation efficiency.

For the multi-stage VIS components, a key requirement is the ability to create a compact, low-frequency pendulum. This is exemplified by devices such as the x-pendulum\cite{XPE}, the inverted pendulum\cite{IPE}, and the Roberts linkage. 

In the Einstein Telescope (ET), one of the planned third-generation gravitational wave telescopes, it is necessary to adopt a more advanced vibration isolation system capable of meeting the sensitivity requirements starting from the 0.1 Hz frequency range. 
Currently, there is a concept to use a low-frequency vibration isolation system based on an inverted pendulum (IP), connecting two stages in series for isolation starting from the 0.1 Hz range. 

However, to address the challenge of independently tuning each stage when stacking two IPs, the development of a different type of low-frequency vibration isolation system is needed. 
The Roberts linkage (RL) has the potential to solve this issue, and a four-point RL has been studied in this context\cite{THR,UWA,MOD,TAM}.

The three-point RL\cite{MTM}, as depicted in FIG.~\ref{fig:01RLconfig}, is distinguished by:
\begin{itemize}
\item 
Its ability, similar to four-point RL, to maintain near-horizontal center of mass (COM) movements allows for fine-tuning of the resonance frequency through minor adjustments to the COM position. With a well-designed, concave operational plane, it enables a compact, long-period pendulum with minimal material property impact, relying solely on its geometric design for performance. 
\item 
In setups with more than three suspension wires, typically only three actively support the rigid body at any given moment. As the pendulum moves, the supporting wires may shift, leading to nonlinear behavior and instability. In contrast, the three-point Roberts linkage avoids over-constraining by ensuring that all suspension wires remain consistently engaged, providing stable and reliable operation.
\end{itemize}

Therefore, it is essential to theoretically investigate the characteristics required for the design and operation of the three-point Roberts linkage.

\section{Three-Point Roberts linkage}
\begin{figure}
\includegraphics[width=\columnwidth]{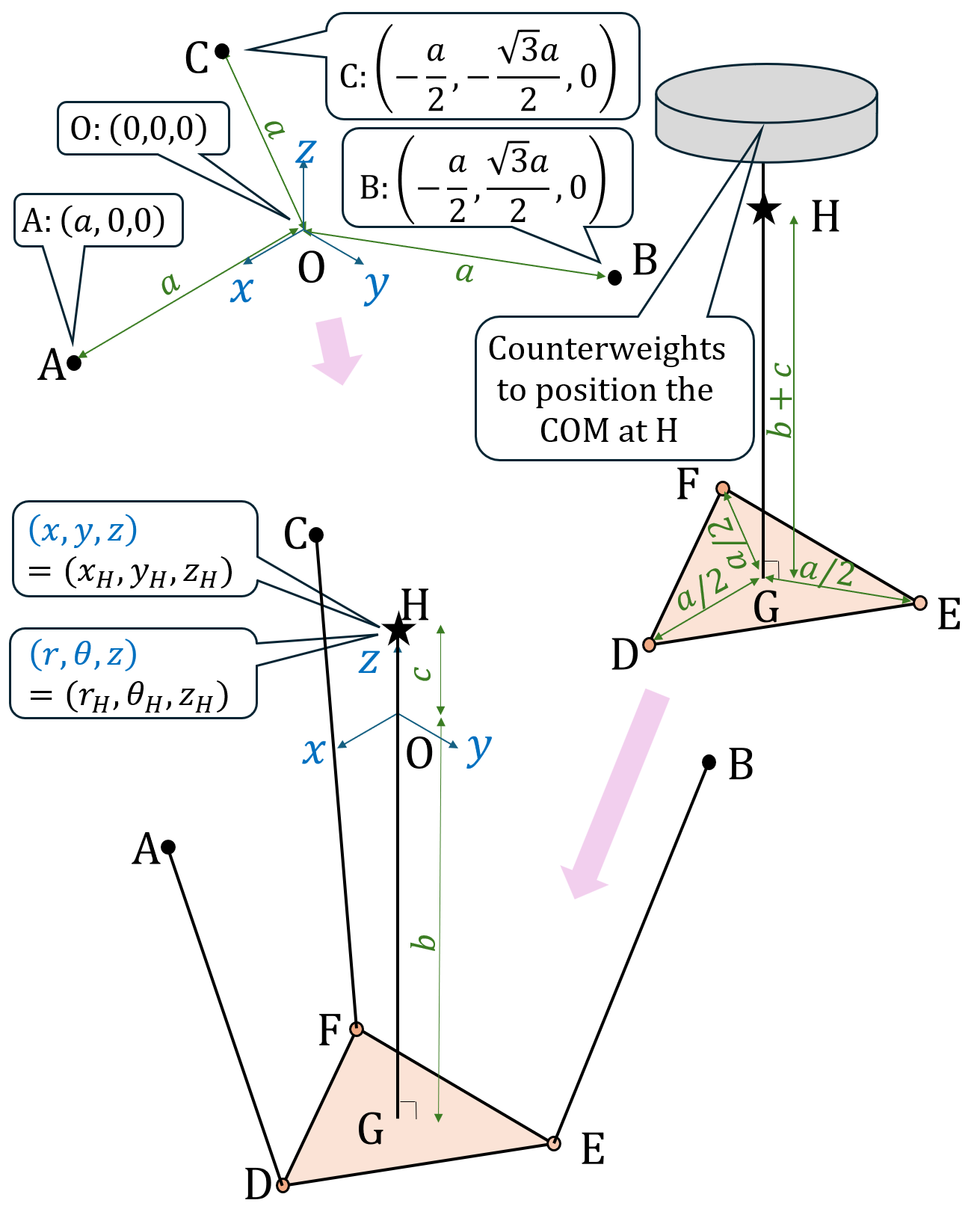}
\caption{\label{fig:01RLconfig}
Three-point Roberts Linkage (RL): 
The top left image shows the system at rest, along with the coordinate setup. 
The origin O is at the centroid of an equilateral suspension triangle in the horizontal plane, with suspension points A, B, and C at $(a,0,0)$, $(-a/2, \sqrt{3}a/2,0)$, and $(-a/2, -\sqrt{3}a/2,0)$, respectively.
One end of each wire, with an equal length of $\sqrt{(a/2)^2+b^2}$, is connected to points A, B, and C, while the other end is connected to points D, E, and F on the rigid body, as shown in the top right image. 
The rigid body is equipped with counterweights to position the center of mass (COM) at the central axis, at a height of $2b+c$ from the base. 
The bottom image shows the position where the system is suspended with the base horizontal and at rest. 
In this position, the center of mass is located at $(0,0,c)$, and the base of the rigid body lies in the $z=-b$ plane. 
Both Cartesian coordinates $(x_H,y_H,z_H)$ and cylindrical coordinates $(r_H,\theta_H,z_H)$ are used to represent the position of the COM, H.}
\end{figure}

This article examines a three-point Roberts linkage (RL) as depicted in FIG.~\ref{fig:01RLconfig}. 
The origin O is placed at the centroid of the `suspension triangle' $\triangle$ABC, which is an equilateral triangle lying in a horizontal plane, with the x-axis pointing towards `suspension point' A and the z-axis directed upwards. The suspension points, A, B, and C, all lie in the xy-plane and are positioned at A\((a,0,0) \),  B\((-a/2,\sqrt{3}a/2,0) \), and C\( (-a/2,-\sqrt{3}a/2,0) \). 
Each wire, fixed at one end to the suspension points A, B, and C, has an equal length of $\sqrt{(a/2)^2+b^2}$, with the other ends attached to points D, E, and F on the rigid body, suspending it.
Points D, E, and F form an equilateral triangle on the base of the rigid body, located at a distance of $a/2$ from the central axis, and referred to as the `fixed points.'
The center of mass (COM) of the rigid body is located at point H, which lies on the central axis at a distance of $b+c$ above point G, where the axis intersects the `base triangle' $\triangle$DEF.

Although counterweights (the gray area in FIG.~\ref{fig:01RLconfig}) are attached to the upper part of the rigid body to position the COM at H, and their effects on the COM position, body mass, and moment of inertia are considered, their shape does not directly influence the system’s behavior, and thus they are omitted from the subsequent discussion.
The wires at both the fixed and suspension points are free to rotate without restriction. 
In its resting state, with the base triangle parallel to the suspension triangle (the `horizontal position'), the base triangle is positioned $b$ units below the suspension triangle, and the COM, denoted as H, is located $c$ units above the suspension plane.

When the rigid body is in the horizontal position, the fixed points are defined at D\((a/2, 0, -b) \),  E\((-a/4, \sqrt{3}a/4, -b) \), and F\((-a/4, -\sqrt{3}a/4, -b) \), with the COM, H located at \( (0, 0, c) \). 

The position of the COM, H during operation is given in Cartesian coordinates as \((x_H, y_H, z_H)\) and in cylindrical coordinates as \((r_H, \theta_H, z_H)\).

In this article, we evaluate the behavior of an RL with specific dimensions. We have constructed and are currently evaluating an RL with dimensions $a$ = 13 cm, $b$ = 40 cm, and $c$ = 0 cm. Henceforth, we refer to the RL with these dimensions as the "prototype dimensions" (PD), and this configuration will be denoted as the PDRL (Prototype Dimensions RL) throughout the article.
Among the PD parameters, $c$ may take values other than 0 cm, but unless otherwise specified, we assume $c$ = 0 cm. The behavioral analysis of this prototype has been reported in a separate article \cite{MTM}.

\begin{figure}
\includegraphics[width=0.6\columnwidth]{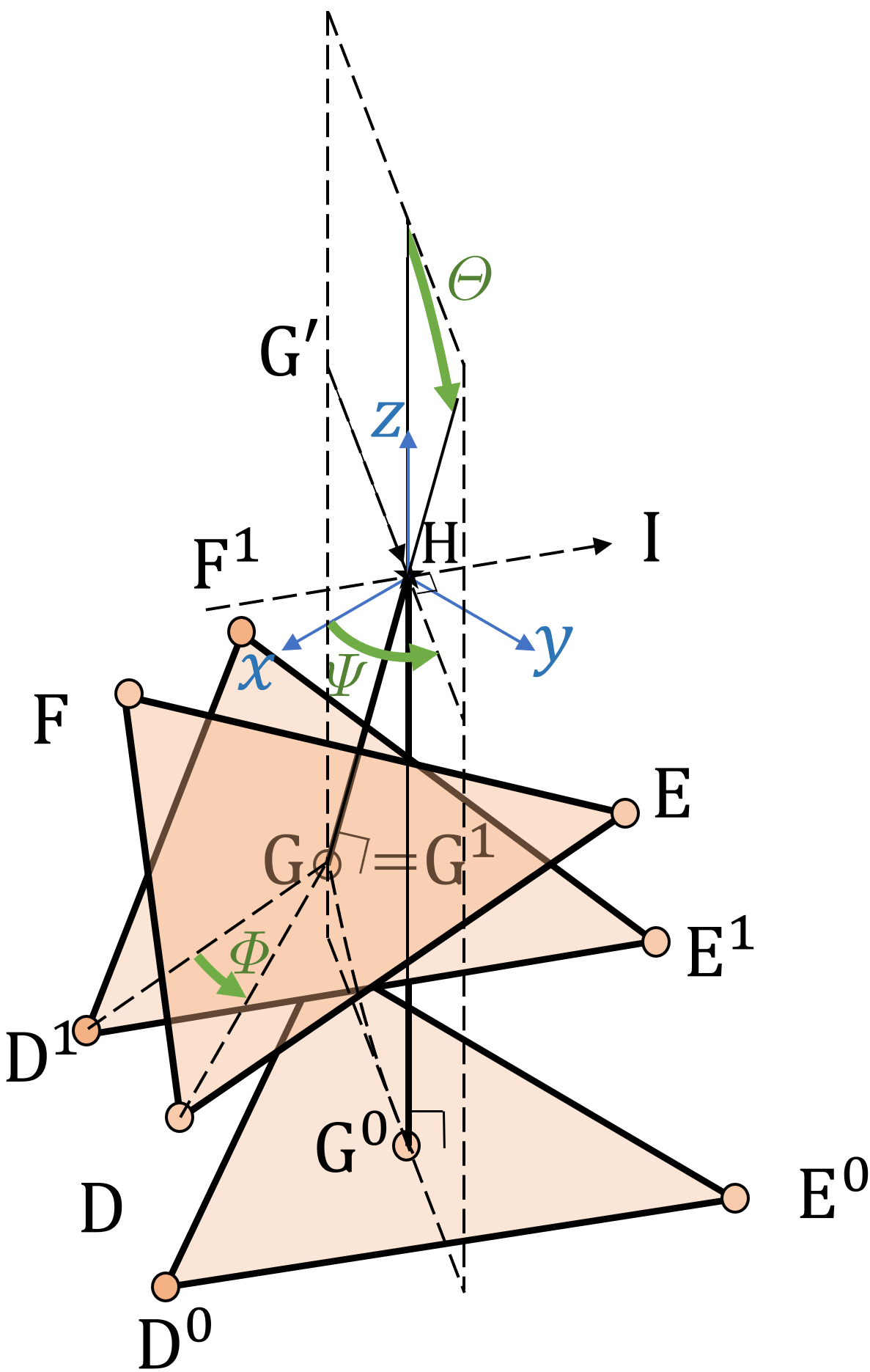}
\caption{\label{fig:02DifAngle} 
Definition of Angles: 
The orientation of the rigid body HDEF is represented by the tilt direction $\varPsi$, tilt angle $\varTheta$, and torsion angle $\varPhi$. $\rm{D^0}$, $\rm{E^0}$, and $\rm{F^0}$ are the positions of D, E, and F when the rigid body is in its nominal rest position with the base triangle horizontal. The angle $\varPsi$ is the angle between the projection of the central axis GH onto the x-y plane (denoted as G'H) and the x-axis. This angle indicates the direction of tilting but not the rotation angle of the suspension. The tilt angle $\varTheta$ is the angle between the central axis GH and the z-axis. After the rigid body tilts by $\varTheta$ in the $\varPsi$ direction (rotating around HI by $\varTheta$), the corner points shift to $\rm{D^1}$, $\rm{E^1}$, and $\rm{F^1}$. In this tilted position, $\varPhi$ represents the torsion angle, indicating the rotation around the central axis GH from $\rm{D^1E^1F^1}$ to DEF.
}
\end{figure}

In this article, the rotational position of the rigid body during the pendulum-like motion of the RL is described using three angles: tilt direction $\varPsi$, tilt angle $\varTheta$, and torsion angle $\varPhi$.
As shown in FIG.~\ref{fig:02DifAngle}, the direction in which the rigid body axis GH tilts is represented by the direction $\varPsi$, and the angle $\varTheta$. 
In the coordinate system where the origin is translated to the center of mass H, $\varPsi$ is the angle in the xy-plane measured from the x-axis to $\vec{\rm{G'H}}$, which is the projection of the vector $\vec{\rm{GH}}$ onto the xy-plane.
The tilt angle $\varTheta$ is the angle between the rigid body axis GH and the z-axis. 
In other words, a tilt of direction $\varPsi$ and angle $\varTheta$ is equivalent to a clockwise rotation of $\varTheta$ around the HI axis, which is at an angle of $\varPsi+\pi/2$ from the x-axis in the xy-plane. 
$\varPhi$ represents the rotation around the axis GH. 

As shown in FIG.~\ref{fig:02DifAngle}, a rigid body DEFH with tilt direction $\varPsi$, tilt angle $\varTheta$, and torsion angle $\varPhi$ is the result of rotating the rigid body from its horizontal position $\rm{D_0E_0F_0H}$, tilted by $\varTheta$ in the direction $\varPsi$ to the position $\rm{D_1E_1F_1H}$, and then rotating it by $\varPhi$ around the rigid body axis.

\begin{figure}
\includegraphics[width=0.25\columnwidth]{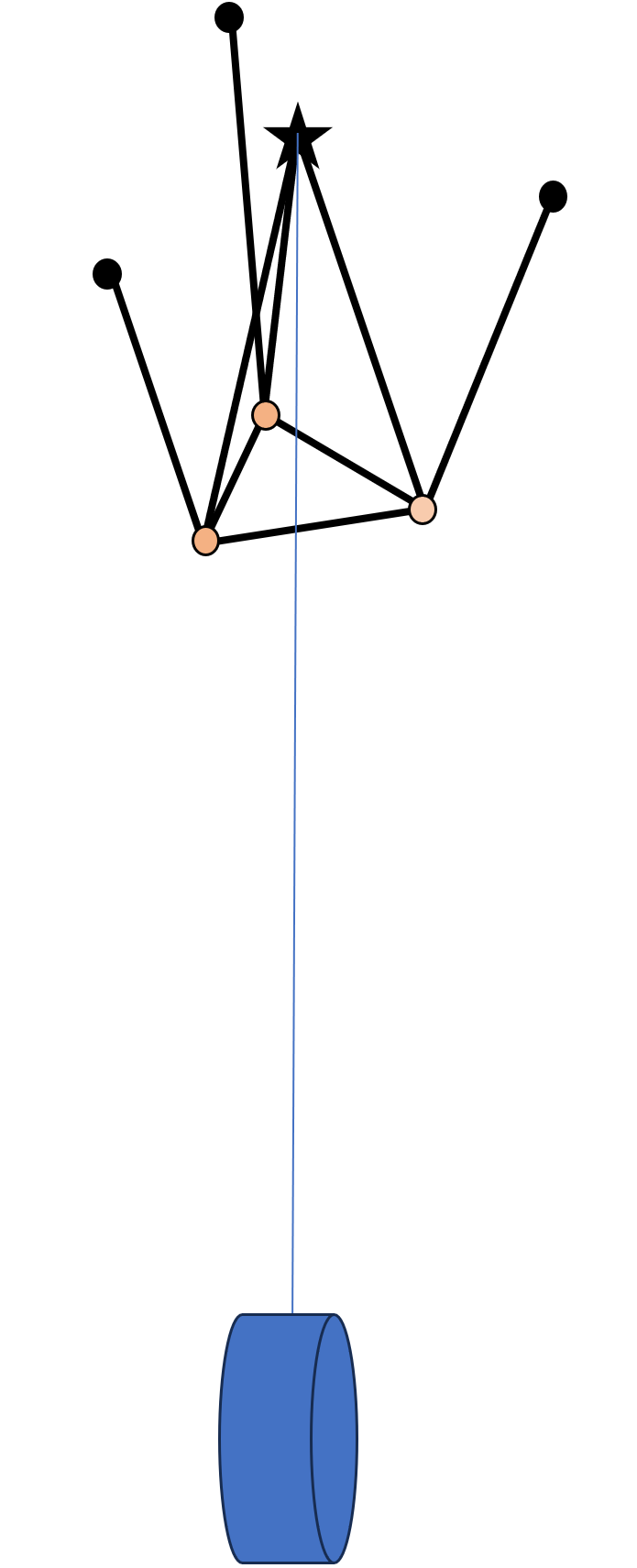}
\caption{\label{fig:20practical} 
The Intended Application:
The next stage of the VIS, such as a simple pendulum, is expected to be suspended at the COM of the RL.
}
\end{figure}

The intended application of the RL discussed in this article is shown in FIG.~\ref{fig:20practical}. We envision incorporating it into the VIS for the test mass in gravitational wave telescopes. As illustrated in the figure, the next stage of the VIS, such as a simple pendulum, is expected to be suspended at the COM of the RL. In the multi-stage vibration isolation system, vibration isolation in the vertical and torsional directions is managed by separate systems, while the RL is responsible solely for isolating horizontal translational motion. The characteristics under this usage can be described by the Lagrangian when the COM moves horizontally relative to the suspension point. 
Therefore, in the coordinate system where the suspension point is stationary, the RL's characteristics are evaluated by determining the vertical position of the COM, which defines the potential energy, and the rotation of the RL, which defines the kinetic energy, both as functions of the horizontal position of the COM.

\section{Rigid Body Motion Analysis Model}
\begin{figure}
\includegraphics[width=1\columnwidth]{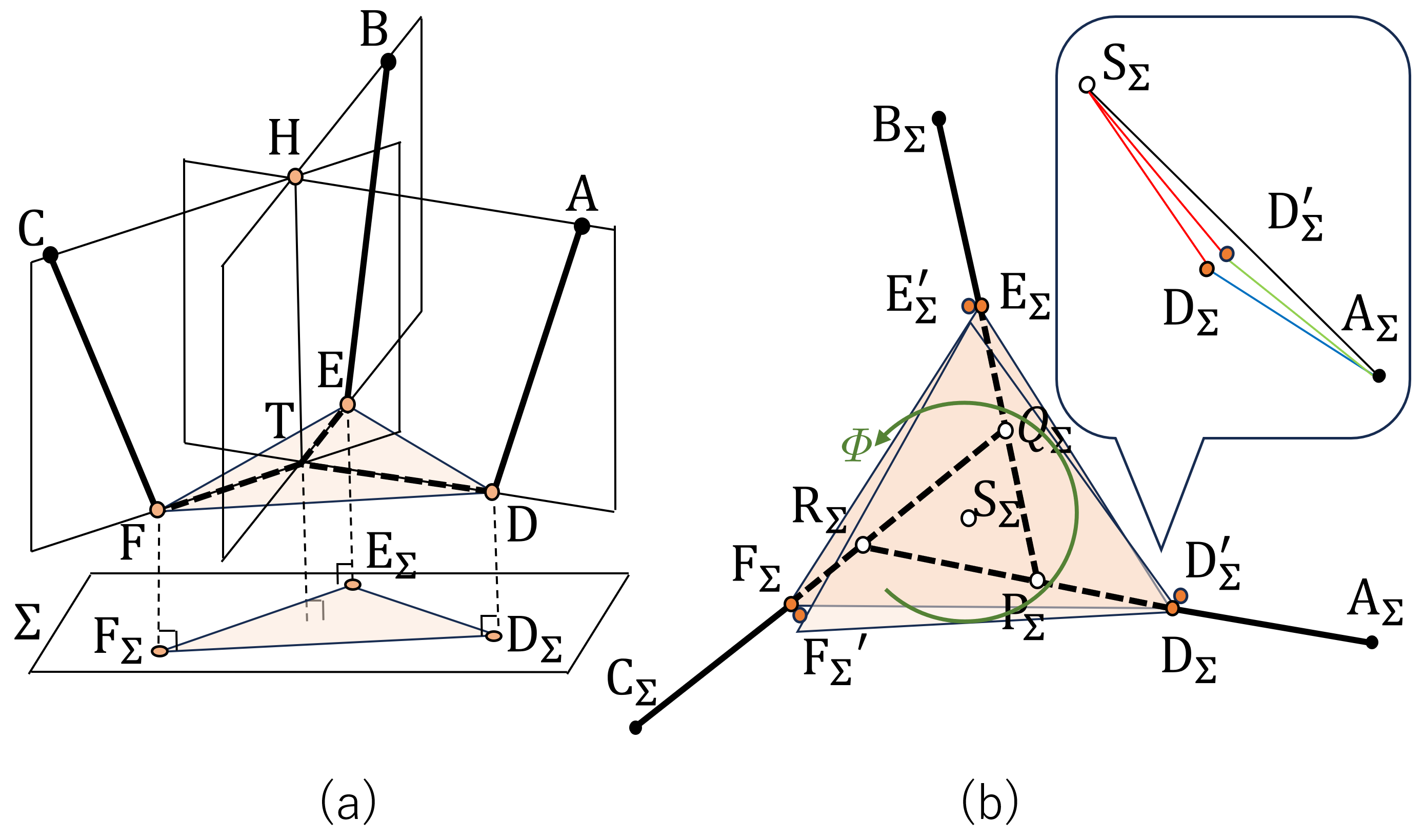}
\caption{\label{fig:04Relux}
Constraints for Minimizing $z_H$ Under Fixed $x_H$ and $y_H$: 
Panel (a) shows the configuration where the three planes, each formed by one of the three wires and the COM, H, intersect along a single line HT, achieving the minimum $z_H$. 
Panel (b) illustrates that the three planes do not intersect along a line but converge at a single point H, forming a triangular pyramid HPQR. 
This configuration is projected onto plane $\Sigma$.
Here, a minor rotation from DEF to D'E'F' around the axis HS, as indicated in (b), allows the wires to slacken, which can lower the body's position, thus demonstrating that (b) does not achieve the minimum  $z_H$ position.
}
\end{figure}

Rigid body motion includes both translational and rotational aspects, featuring six degrees of freedom. For the suspended rigid body, three constraints stipulate that the lengths of the wires AD, BE, and CF are fixed, thereby reducing the available degrees of freedom to three. Consequently, even when the horizontal positions of the center of mass (COM), namely $x_H$ and $y_H$, are specified, one degree of freedom remains, leaving the height $z_H$ undetermined. This remaining degree of freedom is associated with the torsion angle $\varPhi$, which will be discussed further using FIG.~\ref{fig:04Relux}. Due to this degree of freedom, $z_H$ can vary, and under certain conditions, it can reach a minimal value. Thus, during normal operation of the linkage, it is assumed that this minimum value is achieved. The basis for this assumption in the evaluation of vibration isolation system behavior is discussed further in the `Discussion' section.

Previous studies\cite{PE} have utilized the method of Lagrange multipliers to find the minimum value of $z_H$. In contrast, this article employs a geometric approach to this determination. 
Geometrically, the constraint for achieving the minimum $z_H$ requires that the three planes, each formed by one of the wires and the COM, H, converge along a single line HT, as depicted in FIG.~\ref{fig:04Relux}(a).
Conversely, when the three planes do not intersect along a single line and instead converge at a single point, H, they form a triangular pyramid HPQR along with the base face DEF of the rigid body. This pyramid is then projected onto a plane $\Sigma$, which is perpendicular to the rotational axis HS, with S being a point within triangle PQR. 
This projection is depicted in panel (b), as indicated by the subscript $\Sigma$, which denotes projections onto the $\Sigma$ plane.
From the torque-experiencing configuration of the suspended DEF, it can be inferred that a minor torsion-like rotation of the rigid body from DEF to D'E'F' around the axis HS, centered at $S_\Sigma$ in (b), permits the wires to slacken, thereby lowering the body's position. In contrast, the configuration illustrated in (a) does not allow any minor rotation that would result in a lower position, thus confirming it as the minimal value of $z_H$

Imposing this constraint, along with the three constraints for constant wire lengths, limits the degrees of freedom of the suspended rigid body to two. Therefore, by specifying the horizontal positions $x_H$ and $y_H$ of the COM, the position and orientation of the rigid body are uniquely determined.

Given these constraints, general analytical solutions are challenging to derive, requiring the use of numerical methods for practical resolution. This study employs Newton's method to find numerical solutions, which are outlined as follows: 
\begin{itemize} 
\item Variables (12 in total): These include the positions of four points on the rigid body $(x_D, y_D, z_D)$, $(x_E, y_E, z_E)$, $(x_F, y_F, z_F)$, and $(x_H, y_H, z_H)$. 
\item Constraints (10 in total): 
\begin{itemize} 
\item Constant distances for AD, BE, and CF (3 constraints), 
\item Intersection of planes ADH, BEH, and CFH along a single line (1 constraint), 
\item Constant distances for DE, EF, and FD (3 constraints), 
\item Constant distances for HD, HE, and HF (3 constraints). 
\end{itemize} 
\end{itemize}

By imposing these constraints, the positions of the four points are calculated as functions of $x_H$ and $y_H$, allowing for the determination of $z_H$, $\varPsi$, $\varTheta$, and $\varPhi$ as functions of either $(x_H, y_H)$ or $(r_H, \theta_H)$.  
Calculations were performed using Mathematica's FindRoot function with 64-digit precision.

\section{Essential Parameters for VIS and Derivation of Formulas for These}
The goal of this chapter is to derive formulas for calculating essential parameters for RL when utilized as a Vibration Isolation System (VIS), based on its size parameters. Initially, Section A outlines the parameters necessary for determining equivalent pendulum lengths and assessing the contributions of the rigid body's moment of inertia, which are crucial for its effectiveness as a VIS. Ideally, if one could analytically determine the position and orientation of the rigid body from specified size parameters and the horizontal position of the COM, these parameters would be straightforwardly defined. However, this is challenging in practice. Consequently, this article initially focuses on two special cases to facilitate a deeper understanding of the system's behavior. In Section B, we present analytical solutions for the operation of RL along its plane of symmetry for arbitrary sizes. Section C details the numerical analysis of a PDRL, exploring its behavior in various directions. Building upon these findings, Section D formulates hypotheses concerning the behavior of RL in arbitrary directions and sizes, and verifies these hypotheses across different dimensions to validate the general applicability of the derived solutions.

\subsection{\label{sec:level2}Essential Parameters for Motion Analysis}
To perform the motion analysis required for utilizing the Roberts linkage (RL) as a horizontal vibration isolation system in gravitational wave telescopes, it is sufficient to describe the potential and kinetic energy in the neighborhood of the stable equilibrium position, where the COM in the horizontal position $(x_H, y_H)$ is close to $(0, 0)$. 
The potential energy can be determined if the height of the COM, $z_H$, in the vicinity of $(x_H, y_H) = (0, 0)$ is known, which makes the Hessian matrix in this neighborhood essential for deriving the equations of motion.

The rotation of the rigid body can be described by rotations in the $\varTheta$ and $\varPhi$ directions, assuming $\varPsi$ is known. 
If \( d\varTheta/dr_H \) and \( d\varPhi/dr_H \) are provided as the body approaches \( r_H \rightarrow 0 \), the angular velocities in the \( \varTheta \)- and \( \varPhi \)-directions can be derived as functions of \( \dot{x_H} \) and \( \dot{y_H} \), and consequently, the kinetic energy can also be expressed as a function of \( \dot{x_H} \) and \( \dot{y_H} \).

In this chapter, we will derive four types of parameters— the Hessian matrix of $z_H$ with respect to $x_H$ and $y_H$, $\varPsi$, $d\varTheta/dr_H$, and $d\varPhi/dr_H$—specifically focusing on those in the vicinity of the stable equilibrium point. Understanding these parameters enables us to construct the Lagrangian of the system based on the horizontal position and horizontal velocity of the COM near the equilibrium point, thereby facilitating the analysis of its dynamics.

\subsection{\label{sec:level2}2D Motion Analysis Along the Symmetry Plane of the RL}
The RL depicted in FIG.~\ref{fig:01RLconfig} is symmetric with respect to the x-z plane. 
When the COM, H, moves along \(\theta_H = 0\), meaning along the symmetry plane, it maintains symmetry which prevents any twisting and causes the rigid body to tilt along the x-axis, thus,
\begin{eqnarray}
\varPsi=\varPhi=0. 
\label{eq01}
\end{eqnarray}

\begin{figure}
\includegraphics[width=1\columnwidth]{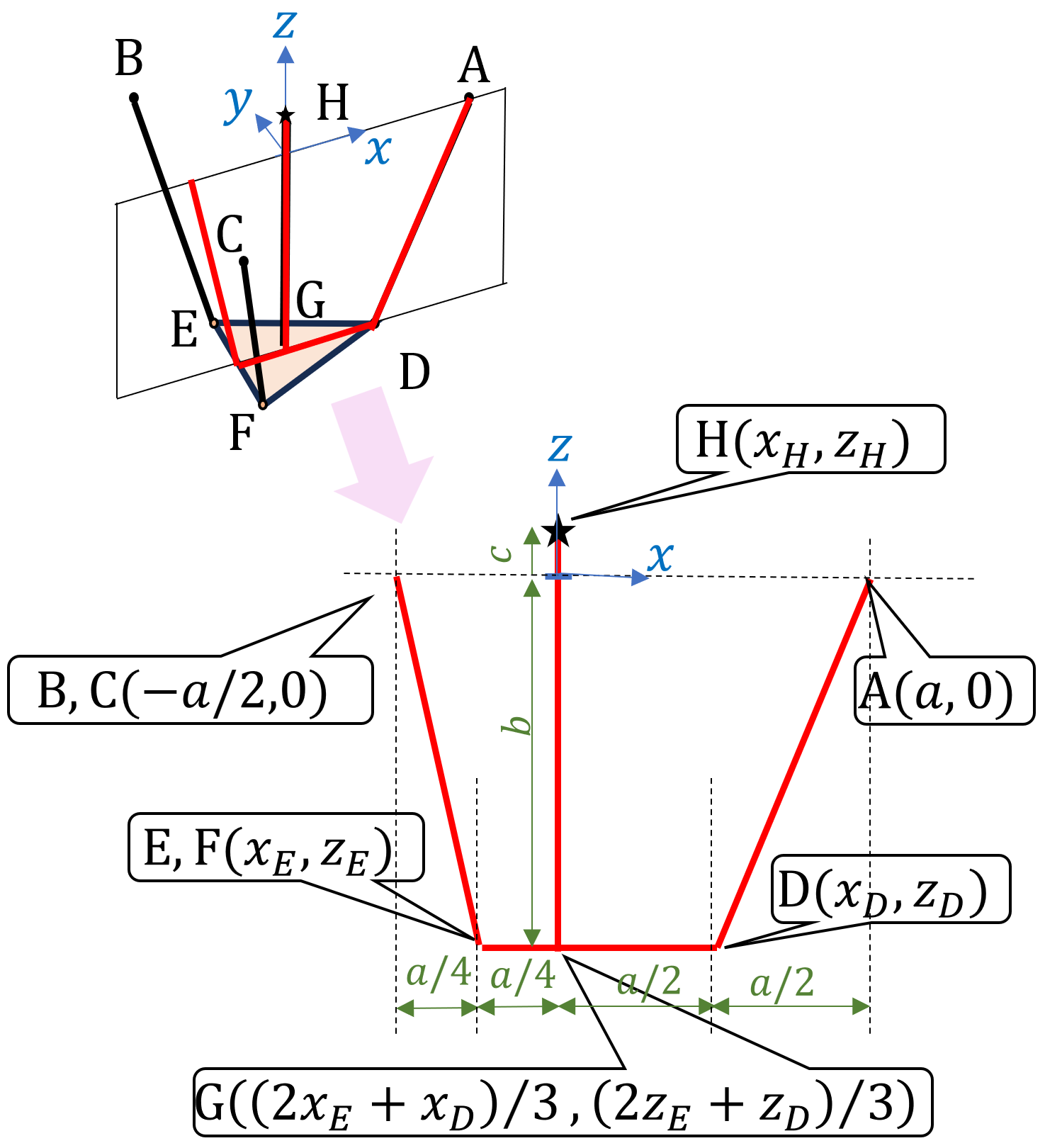}
\caption{\label{fig:133Dto2D}
2D Analysis:
This figure demonstrates that the system can be analyzed in 2D by projecting the COM, H, motion onto the x-z plane. The top left image shows the system as the H moves along the x-z plane, while the bottom right image represents the projection of this system onto the defined plane. The coordinates on this plane are as indicated, facilitating the 2D analysis.}
\end{figure}

Consequently, when the RL is projected onto the x-z plane as illustrated in FIG.~\ref{fig:133Dto2D}, the suspension points B and C, and the fixed points E and F coincide at the same points. The dynamics of the COM \( (x_H, z_H) \) and the tilt angle \( \varTheta \) of the rigid body are represented by a two-dimensional RL marked with a red line in the figure, which can be fully described.

When projected onto the x-z plane, points E and F coincide and are both referred to as point E. 
Describing the system's motion in terms of coordinates $(x_E, z_E)$ for point E, and $(x_D, z_D)$ for point D, the constraints are expressed by three equations corresponding to the constant lengths of line segments AD, BE, and ED. 
Using these three equations to eliminate three variables, and treating H's position specifically as ($x_H$, $y_H$), we can derive $x_H$, $y_H$, $z_E$, $x_D$, and $z_D$ as functions solely dependent on $x_E$. The expressions of $x_H$, $z_H$, and $\varTheta=(z_E-z_D)/(x_D-x_E)$, as functions of $x_E$ are all quite complex. However, from the equations:
\begin{eqnarray}
\frac{d^2 z_H}{d{x_H}^2}
&=& \frac{d}{dx_E}\left(\frac{dz_H}{dx_E}/\frac{dx_H}{dx_E}\right)/\frac{dx_H}{dx_E}
\label{eq02}
\\
\frac{d\varTheta}{dx_H}
&=& \frac{d}{dx_E}\left(\frac{z_E-z_D}{x_D-x_E}\right)/\frac{dx_H}{dx_E}
\label{eq03}
\end{eqnarray}
the expressions for $d^2z/{dx_H}^2$ and $d\varTheta/dx_H$ when $x_H = 0$, that is, when $x_E = -a/4$, are represented by much simpler formulas. Utilizing Mathematica, the following results were obtained.

\begin{eqnarray}
\frac{d^2 z_H}{d{x_H}^2}\bigg\rvert_{x_H=0}&=&\frac{a^2 -4bc}{4b(2b+c)^2}\label{eq06}\\
\label{eq05}\frac{d\varTheta}{dx_H}\bigg\rvert_{x_H=0}&=&\frac{1}{2b+c}
\end{eqnarray}

From  Eq.~(\ref{eq06}), when

\begin{eqnarray}
c=\frac{a^2}{4b}\equiv c_0,\label{eq07}
\end{eqnarray}
 $d^2z/{dx_H}^2\rvert_{x_H=0}$ becomes 0.

\subsection{\label{sec:level2}3D Numerical Motion Analysis of the PDRL}
In this section, we evaluate the behavior of the RL with the prototype dimensions (PD) parameters, where $a = 13$ cm, $b = 40$ cm and $c = 0$ cm, as described earlier. The RL with these dimensions is referred to as the PDRL (Prototype Dimensions RL).
The parameter $c$ is set to 0 cm unless otherwise specified. 
Specifically, we numerically analyze the behavior of the center of mass height $z_H$, tilt direction $\varPsi$, tilt angle $\varTheta$, and torsion angle $\varPhi$ as functions of the horizontal position of the center of mass. 
Particular attention is given to the behavior near the stable point $(x_H,y_H)=(0 \ \rm{cm},0 \ \rm{cm})$ of the essential parameter discussed in Section A.
Moreover, since the behavior of $z_H$ alone is not straightforward when varying the size parameter $c$, we also evaluate the case where $c$ takes a value other than 0 cm.

\subsubsection{\label{sec:level3}Center of Mass Height $\bm z_H$}
\begin{figure}
\includegraphics[width=1\columnwidth]{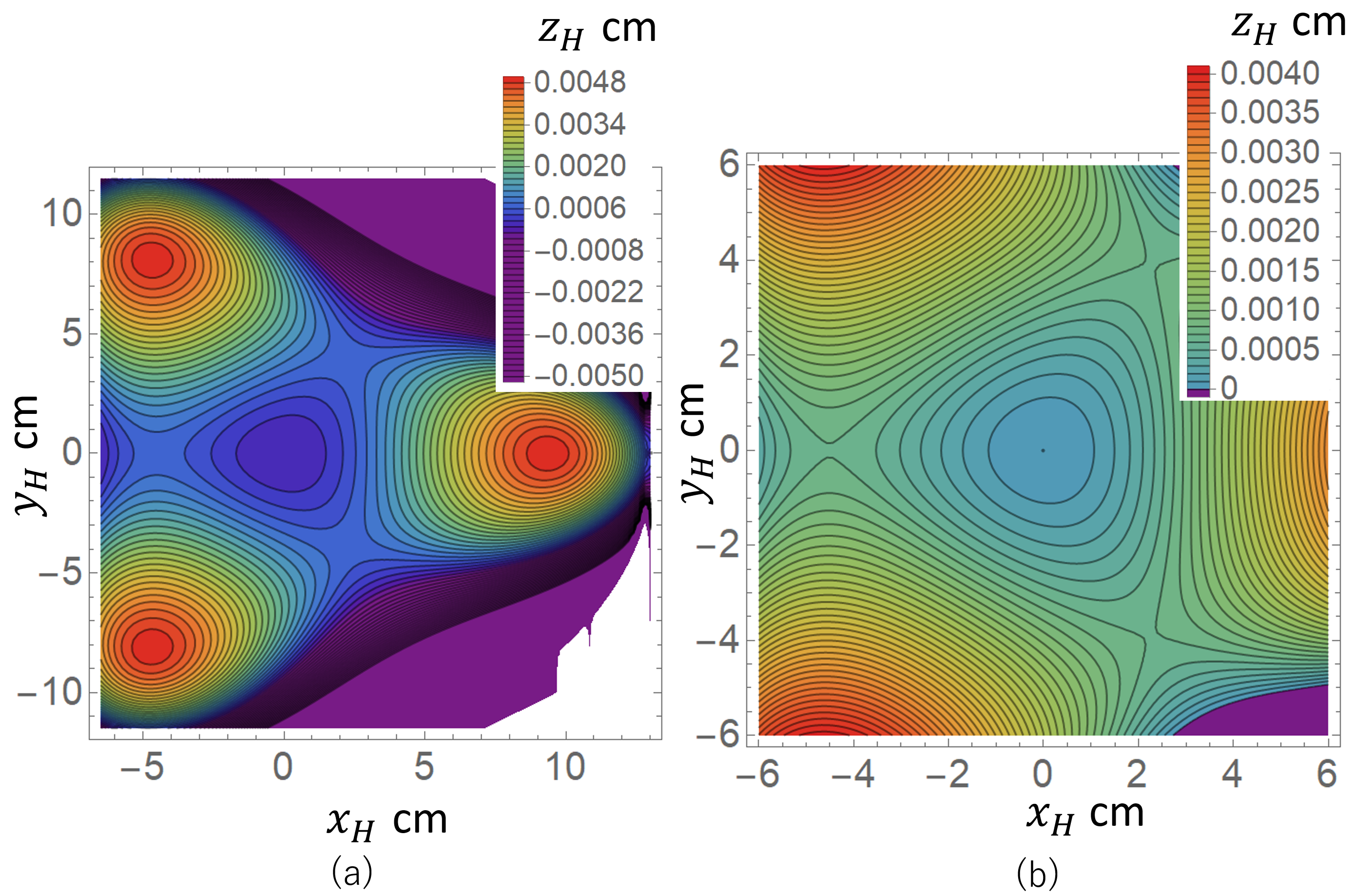}
\caption{\label{fig:05z_contour} 
Contour Maps of the COM's Height: 
This figure illustrates how the COM's height, $z_H$, varies relative to its horizontal position $(x_H, y_H)$ in the PDRL. Panel (a) provides a broad view of the behavior, while panel (b) focuses on the central region. Despite the adjustments, changes in $z_H$ are minimal, approximately 0.005 cm, indicating nearly horizontal motion. The contour lines exhibit three-fold symmetry at larger scales but transition to concentric circles, demonstrating isotropy near the center.}
\end{figure}

\begin{figure}
\includegraphics[width=1\columnwidth]{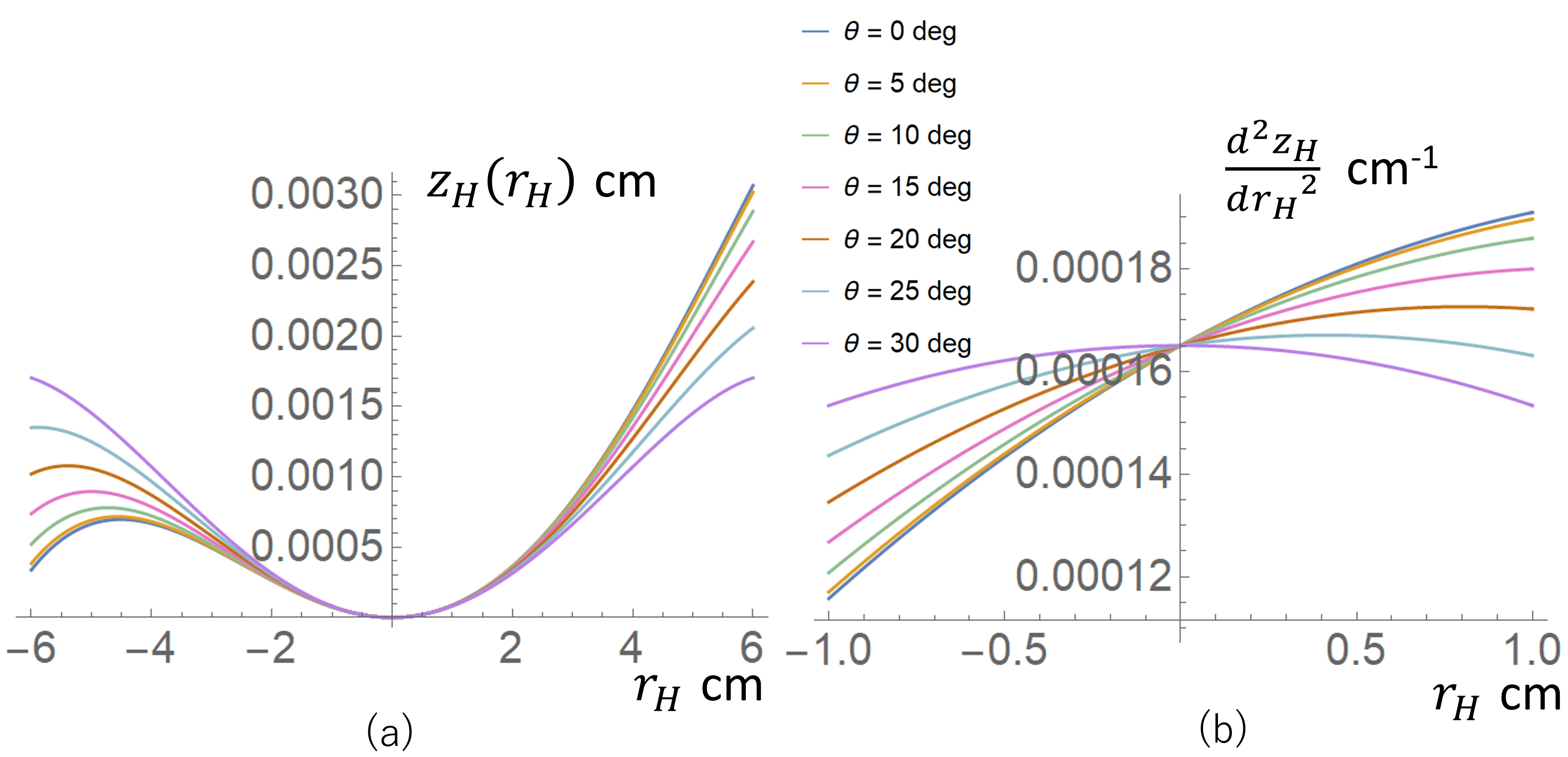}
\caption{\label{fig:06z_r_by_theta} 
The COM's Height and Its Second Derivative: 
Panel (a) of this figure depicts the relationship between the COM's height \( z_H \) and radial distance \( r_H \) within the range \(-6\ \rm{cm} \leq \it{r_H} \leq \rm{6\ cm}\) for the PDRL, illustrating how \( z_H \) varies as a function of \( r_H \) at different orientations \( \theta_H \). Panel (b) focuses on a closer range \(-1\ \rm{cm} \leq \it{r_H}  \leq \rm{1\ cm}\), displaying the second-order derivatives \( d^2z_H/{dr_H}^2 \) as a function of \( r_H \). It emphasizes that these derivatives become isotropic and direction-independent specifically as \( r_H \) approaches zero, with a typical value of approximately \( 1.65 \times 10^{-4}\rm{\ cm^{-1}} \).}
\end{figure}

FIG.~\ref{fig:05z_contour} depicts the behavior of the COM's height $z_H$ as a function of its horizontal position $(x_H, y_H)$ for the PDRL when $c = 0$ cm. 
Cross-sectional views for various directions $\theta_H$ of the COM are presented in FIG.~\ref{fig:06z_r_by_theta}(a). 
Furthermore, FIG.~\ref{fig:06z_r_by_theta}(b) shows $d^2z_H/{dr_H}^2$, the second derivative of $z_H$, which directly relates to the device's natural frequency.
As $\theta_H$ approaches zero, $d^2z_H/{dr_H}^2$ converges to a constant value, irrespective of $\theta_H$. 
This value, 0.000165 $\rm{cm^{-1}}$, aligns with the calculation ${a^2}/{16b^3}$ presented in the previous section's formula (\ref{eq06}).
The equivalent pendulum length is the reciprocal of this, approximately 6060 cm. In other words, a linkage with dimensions of 13 cm × 40 cm can achieve an equivalent pendulum length of 60 m, resulting in a pendulum with a period of about 15 seconds (ignoring the moment of inertia of the rigid body).

\begin{figure}
\includegraphics[width=1\columnwidth]{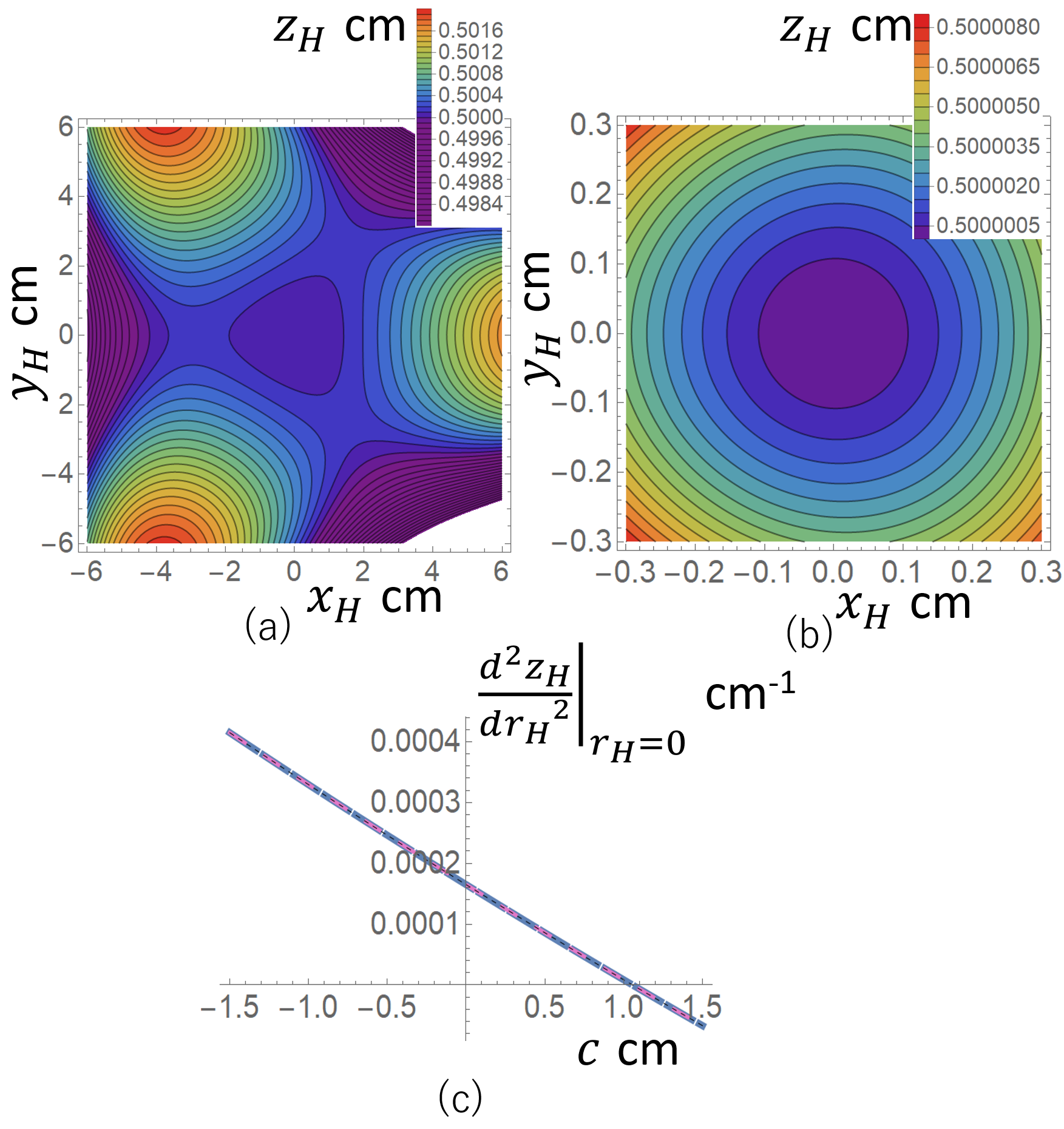}
\caption{\label{fig:08zcontour}
Behavior of \(z_H\) as \(c\) Varies in an RL: 
This figure examines the effects of changing \(c\) in an PDRL. Panel (a) shows a contour map of \(z_H\) for \(c=0.5\) cm, displaying a smaller, shallower central valley at the same scale as FIG.~\ref{fig:05z_contour}(b). Panel (b) zooms into the central region (\(\pm 0.3\)) cm, demonstrating isotropy at \(c=0.5\) cm. Panel (c) charts the variations in the second-order derivative \(d^2z_H/{dr_H}^2\) across different \(c\) values, highlighting a consistent decrease to zero at \(c \approx 1.05\) cm\(\approx a^2/4b\), independent of \( \theta_H \).
}
\end{figure}

Next, a small perturbation is introduced by setting $c=0.5$ cm, which is small relative to the scale of $a=13$ cm and $b=40$ cm, and the behavior under this condition is shown in FIG.~\ref{fig:08zcontour}.
The isotropic region around the minimum value becomes smaller, and the area surrounding the minimum flattens out. 
As depicted in FIG.~\ref{fig:08zcontour}(c), $d^2z_H/{dr_H}^2\rvert_{r_H=0}$ decreases uniformly with $\theta_H$ as $c$ increases, reaching zero when $c\approx1.05$ cm.
This value, calculated as $a^2/4b\approx1.05$ cm, is consistent with the formula (\ref{eq07}) derived in the previous section.

The analysis confirms that, specifically in this case, the second derivative of $z_H$ with respect to $r_H$ at $r_H = 0$ remains constant for any value of $\theta_H$, consistent with the value calculated in Equation (\ref{eq06}) from the previous section.
\begin{equation}
\left.\frac{d^2 z_H}{d r_H^2}\right|_{r_H=0} = \text{constant}, \quad \text{for all } \theta_H.
\label{eq08}
\end{equation}

\subsubsection{\label{sec:level3}Tilt Direction $\varPsi$}
\begin{figure}
\includegraphics[width=1\columnwidth]{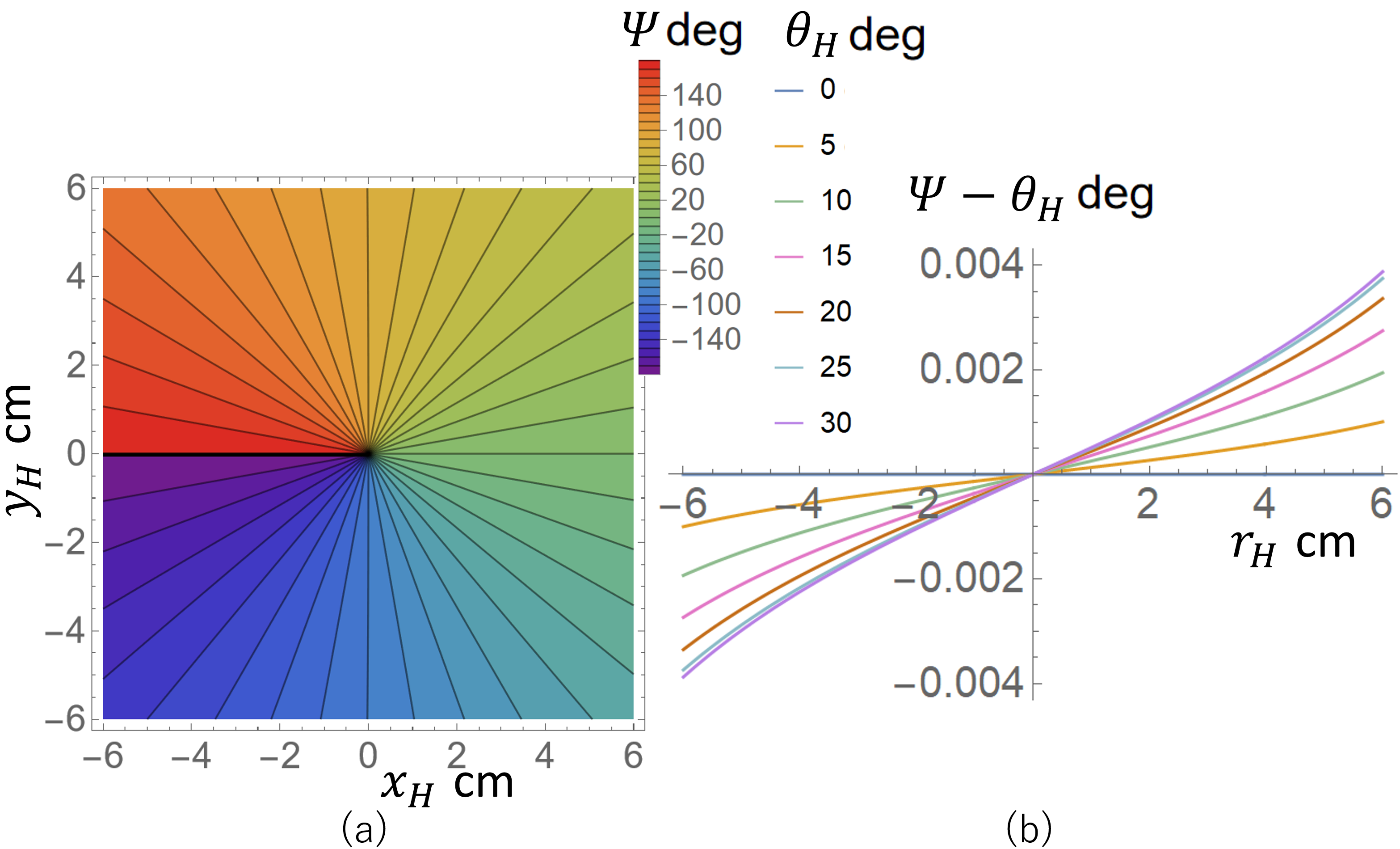}
\caption{\label{fig:10Psi_by_theta}
Behavior of the Tilt Direction $\varPsi$ for the PDRL: 
(a) presents a contour map showing $\varPsi$ in relation to the horizontal position of the COM \( (x_H, y_H) \), where \( (x_H, y_H) = r_H(\cos \theta_H, \sin \theta_H) \). 
(b) explores the dependence of $\varPsi$ on the radial distance \( r_H \) from the origin and its deviation from \( \theta_H \) for various values of \( \theta_H \). It is observed that $\varPsi$ closely aligns with \( \theta_H \) as \( r_H \) approaches zero, demonstrating minimal radial displacements.}
\end{figure}

The tilt direction $\varPsi$ determines the direction in which the system tilts. 
In other words, it indicates which direction's moment of inertia contributes to the system’s kinetic energy. 
In the PDRL, $\varPsi$ is shown as a function of the COM's horizontal position $(x_H, y_H)$ in FIG.~\ref{fig:10Psi_by_theta}(a).
From the figure, it can be observed that $\theta_H$ and $\varPsi$ are approximately equal, meaning that the rigid body tends to tilt in the direction of the COM's movement. 
As discussed in Section B, $\varPsi\equiv\theta_H$ holds true for motion along the symmetry plane, such as the $x-z$ plane. 
To verify this relationship at other angles, the deviation between $\varPsi$ and $\theta_H$ as a function of $r_H(=\sqrt{x_H^2+y_H^2})$ is plotted in FIG.~\ref{fig:10Psi_by_theta}(b) for various $\theta_H$ values. 
Although $\varPsi \ne \theta_H$ for angles other than the symmetry plane ($\theta_H=0$), the figure confirms that near "$r_H=0$", $\theta_H=0$, implying that the equation 
\begin{eqnarray}
    \lim_{r_H \to 0} \varPsi = \theta_H, \quad \text{for all } \theta_H
    \label{eq09}
\end{eqnarray}
holds.

\subsubsection{\label{sec:level3}Tilt Angle $\varTheta$}
\begin{figure}
\includegraphics[width=1\columnwidth]{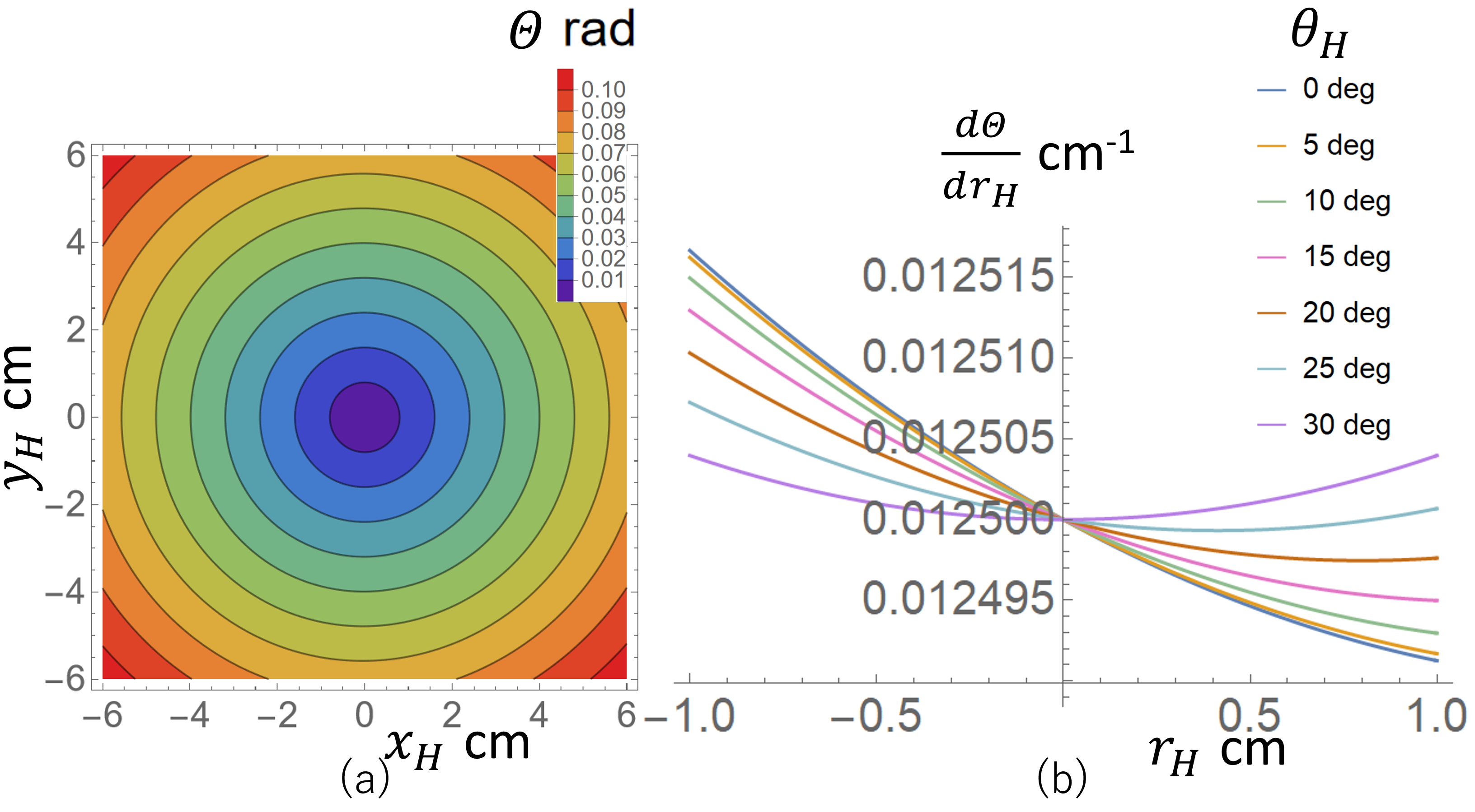}
\caption{\label{fig:11Theta}
Behavior of the Tilt Angle \( \varTheta \) for the PDRL: 
(a) presents a contour map showing \( \varTheta \) in relation to the horizontal position of the COM \( (x_H, y_H) \), where \( (x_H, y_H) = r_H(\cos \theta_H, \sin \theta_H) \). 
(b) explores how \( d\varTheta/dr_H \), the rate of change of \( \varTheta \) with respect to \( r_H \), varies across different \( \theta_H \) values. It is noted that \( d\varTheta/dr_H \) converges to a constant value of 0.0125 $\rm{cm^{-1}}$ at minimal radial displacements, indicating a significant consistency in tilt behavior as \( r_H \) approaches zero.}
\end{figure}

The tilt angle $\varTheta$ represents how much the rigid body tilts in the $\varPsi$ direction. 
In other words, it determines the contribution of the tilt-direction moment of inertia when calculating the kinetic energy from the horizontal movement of the COM. 
For the PDRL, $\varTheta$ is shown as a function of the horizontal position of the COM $(x_H, y_H)$ in FIG.~\ref{fig:11Theta}(a). From the fact that the contour lines are nearly circular and evenly spaced, it can be inferred that $\varTheta$ is roughly proportional to $r_H$. 
As the response to small vibrations near the origin is determined by $d\varTheta/dr_H$, as discussed in Section A, we calculate this as a function of $r_H$ for various $\theta_H$ values and present it in FIG.~\ref{fig:11Theta}(b). 
Although the shape of the graph changes with different $\theta_H$ values, it converges to a single value as $r_H$ approaches zero. 
That is, 

\begin{eqnarray}
\lim_{r_H \to 0} \frac{d\varTheta}{dr_H} = \text{constant,} \quad \text{for all } \theta_H.
\label{eq10}
\end{eqnarray}

The constant value 0.0125 $\rm{cm^{-1}}$ for the PDRL is consistent with Equation (\ref{eq05}) in Section B.

\subsubsection{\label{sec:level3}Torsion Angle $\varPhi$}
The torsion angle $\varPhi$ represents the rotation around the rigid body axis. 
Even when $(x_H,y_H)$, the horizontal position of the COM, is specified, the RL has a degree of freedom for motion in the $\varPhi$ direction, allowing various values of $\varPhi$. However, as described in Chapter III, this article assumes the value of $\varPhi$ that minimizes the vertical position of the COM $(z_H)$. Thus, $\varPhi$ is uniquely determined by $(x_H,y_H)$, and the analysis is conducted based on this assumption.

\begin{figure}
\includegraphics[width=1\columnwidth]{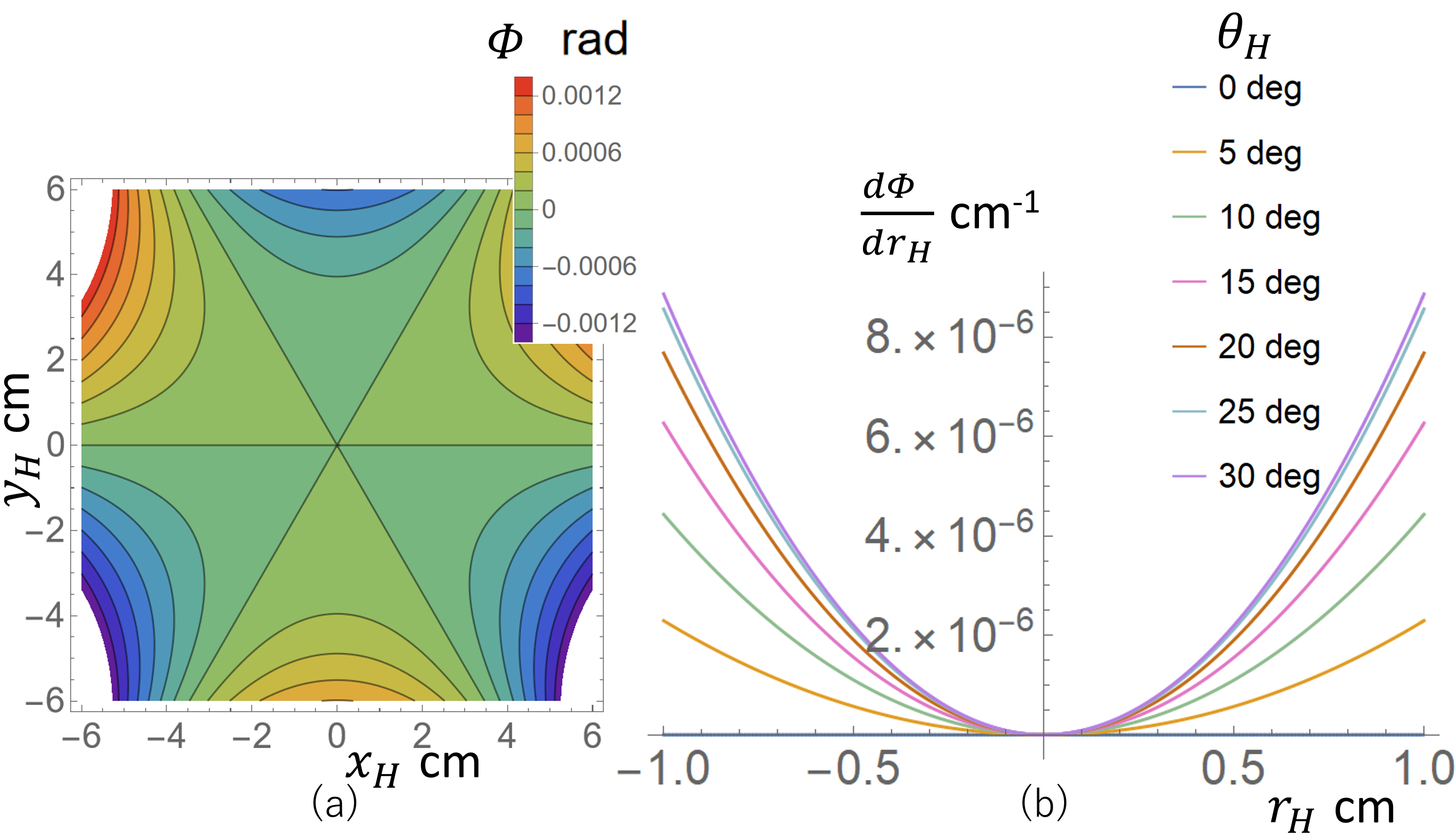}
\caption{\label{fig:12Phi}
Behavior of the Torsion Angle \( \varPhi \) for the PDRL:
(a) presents a contour map showing \( \varPhi \) relative to the horizontal position of the COM \( (x_H, y_H) \), where \( (x_H, y_H) = r_H(\cos \theta_H, \sin \theta_H) \).
(b) explores the relationship between the radial distance \( r_H \) from the COM and the rate of change \( d\varPhi/dr_H \) across various \( \theta_H \) values. 
It is observed that \( d\varPhi/dr_H \) tends towards zero as \( r_H \) approaches zero.}
\end{figure}

For the PDRL, $\varPhi$ is shown as a function of the center of mass's horizontal position $(x_H, y_H)$ in FIG.~\ref{fig:12Phi}(a). 
As mentioned in Section B, for motion along the symmetry planes ($\theta_H = n \pi/3$, $n$: integer), $\varPhi$ remains identically zero. 
However, for other angular directions, it can be observed that not only tilt $\varTheta$ but also torsion $\varPhi$ arises as the center of mass is displaced horizontally. 
Similarly to $\varTheta$, $d\varPhi/dr_H$ is calculated as a function of $r_H$ for various values of $\theta_H$, and the results are presented in FIG.~\ref{fig:12Phi}(b). 
As the deviation from the symmetry axis increases, $d\varPhi/dr_H$ deviates from zero, but even in such cases, as $r_H$ approaches zero, $d\varPhi/dr_H$ converges to zero. That is,

\begin{eqnarray}
\lim_{r_H \to 0} \frac{d\varPhi}{dr_H} = 0, \quad \text{for all } \theta_H.
\label{eq11}
\end{eqnarray}

\subsection{\label{sec:level2}Hypothesis and Verification}
Up to this point, our discussion had spanned the analysis of directional behavior for Roberts linkages of arbitrary dimensions and the evaluation of behavior in arbitrary directions for specific-sized RLs. Given the consistent findings and absence of discrepancies between these analyses, we were then poised to formulate hypotheses regarding the behavior of RLs across any dimensions and directions, and proceeded to systematically evaluate these hypotheses.

\subsubsection{\label{sec:level3}Formulating Hypotheses for RL Behavior}
From Equations (\ref{eq06}) and (\ref{eq08}), it is posited for RLs of any dimension that:
\begin{eqnarray}
    \frac{d^2 z_H}{d{r_H}^2}\bigg\rvert_{r_H=0}=\frac{a^2 -4bc}{4b(2b+c)^2}, \quad \text{for all } \theta_H.
    \label{eq12}
\end{eqnarray}
This can also be expressed using the Hessian matrix as follows:
\begin{eqnarray}
       \left( \begin{array}{cc} \frac{\partial^2 z_H}{\partial{x_H}^2} & \frac{\partial^2 z_H}{\partial x_H\partial y_H} \\ \frac{\partial^2 z_H}{\partial x_H\partial y_H} & \frac{\partial^2 z_H}{\partial{y_H}^2} \end{array} \right) \bigg\rvert_{r_H=0}= \frac{a^2 - 4bc}{4b(2b + c)^2} \left( \begin{array}{cc} 1 & 0 \\ 0 & 1 \end{array} \right).
    \label{eq13}
\end{eqnarray}
Similarly, from Equations (\ref{eq01}) and (\ref{eq09}), (\ref{eq05}) and (\ref{eq10}), and (\ref{eq01}) and (\ref{eq11}), it is presumed for RLs of any dimension that:
\begin{eqnarray}
    &&\lim_{r_H \to 0} \varPsi = \theta_H, \quad \text{for all } \theta_H.
    \label{eq14}\\
    &&\lim_{r_H \to 0} \frac{d\varTheta}{dr_H} = \frac{1}{2b+c}, \quad \text{for all } \theta_H.
    \label{eq15}\\
    &&\lim_{r_H \to 0} \frac{d\varPhi}{dr_H} = 0, \quad \text{for all } \theta_H.
    \label{eq16}
\end{eqnarray}
These relationships, expressed as Equations (\ref{eq13}) to (\ref{eq16}), are posited as hypotheses.

\subsubsection{\label{sec:level3}Verification of Hypotheses}
We conducted numerical verification across various RL dimensions, examining 31 values each for dimensions $a$ and $b$ ranging from 1 to 1000, 11 values for $c$ ranging from 0 up to $c_0$ as defined in Equation (\ref{eq07}), and 7 values for $\theta_H$ from 0 to 30 degrees.

For Equation (\ref{eq13}), we verified that two eigenvalues of the Hessian matrix remained consistent across all 10,571 (31×31×11) combinations of $a, b,$ and $c$.

Assuming the validity of Equations (\ref{eq14}) and (\ref{eq15}), the behavior of the system posits that the rigid body rotates about a fixed point located at $(2b+c)$ units directly beneath the COM. Thus, at the same rigid body position with $c=0$, the horizontal position is given by $({2b}/{2b+c})(x_H ,y_H)$. Consequently, if we denote $\varPsi(x_H,y_H)$ at $c=0$ as $\varPsi_0(x_H,y_H)$, it follows that:
\begin{eqnarray}
\varPsi(x_H,y_H)= \varPsi_0\left(\frac{2b}{2b+c}x_H, \frac{2b}{2b+c}y_H\right)
\end{eqnarray}
This implies that if Equation (\ref{eq14}) is verified at $c=0$, it is expected to hold for any $c$. Similarly, if the validity of Equations (\ref{eq15}) and (\ref{eq16}) is confirmed at $c=0$, they are presumed valid for any $c$. Therefore, verification of these equations at $c=0$ is sufficient. Accordingly, Equations (\ref{eq14}) through (\ref{eq16}) were tested at $c=0$ across 6,727 (31×31×7) combinations of $a, b,$ and $\theta_H$.

These findings affirm that our hypotheses are universally applicable, rendering them suitable for RLs of any dimension in practical applications.

These findings affirm that Equations (\ref{eq13}) to (\ref{eq16}) are universally applicable, rendering them suitable for RLs of any dimension in practical applications.

\section{Influence and Compensation of Structural Variations}
In this chapter, we examined the influence on the COM's gliding plane, represented by $z_H(x_H, y_H)$, when the PDRL was subjected to suspension plane inclination, wire length discrepancies, and horizontal displacement of the COM. The results showed that even small errors significantly affect the rigid body's resting position, and $d^2z_H/{dx_H}^2$ and $d^2z_H/{dy_H}^2$ changes its value anisotropically. Additionally, these three types of errors can potentially cancel each other out to stabilize the rigid body at the center, but issues of anisotropy may still remain.

\subsubsection{\label{sec:level3}Effects of Inclination}
\begin{figure}
\includegraphics[width=1\columnwidth]{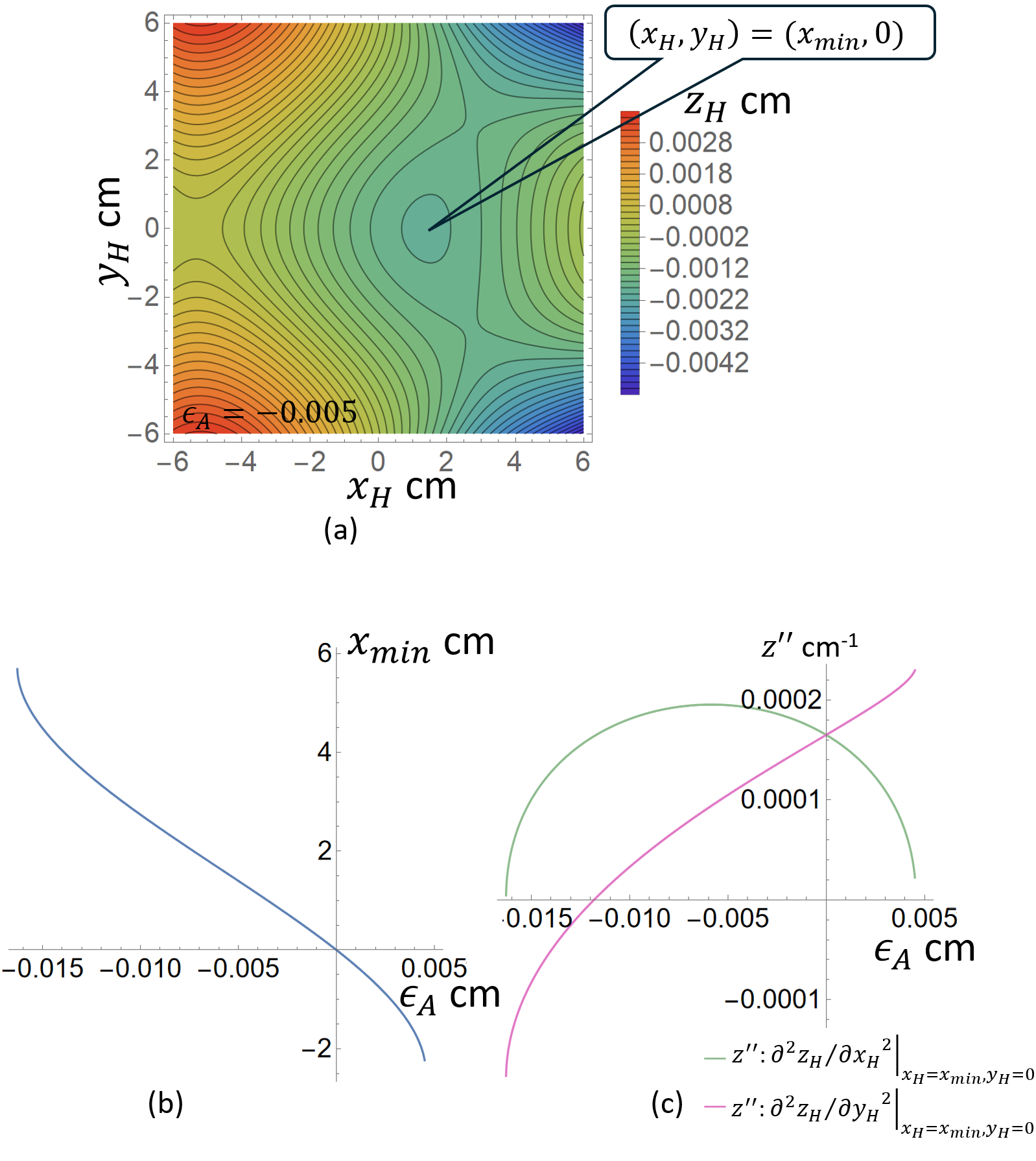}
\caption{\label{fig:15tilt}
Effects of a Minor Tilt Around the Y-Axis: 
This figure explores the effects of tilting on the PDRL by vertically adjusting only suspension point A to positions \( (a, 0, \epsilon_A) \).
Panel (a) displays contour lines of \( z_H(x_H, y_H) \) for \( \epsilon_A = -0.005 \) cm, illustrating the shift of the \( z_H \)'s minimum point along the x-axis and its resulting anisotropic deformation.
Panel (b) shows the relationship between \( \epsilon_A \) and the x-coordinate \( x_{\text{min}} \) of \( z_H \)'s minimum point.
Panel (c) presents how changes in \( \epsilon_A \) affect the second derivatives \( \partial^2 z_H/{\partial x_H}^2 \) and \( \partial^2 z_H/{\partial y_H}^2 \) at the minimum point \( (x_{\text{min}}, 0) \), highlighting the induced anisotropy in the deformation patterns.}
\end{figure}

To investigate the behavior of the PDRL when tilted around the \( y \)-axis, we adjusted the position of suspension point \( A \) from its original location at \( (a,0,0) \) to \( (a,0,\epsilon_A) \) and analyzed its behavior. This analysis included examining the shift in the \( x \)-coordinate of \( z_H \)'s minimum point, denoted as \( x_{\text{min}} \), as illustrated in FIG.~\ref{fig:15tilt}.

FIG.~\ref{fig:15tilt}(a) presents the contour lines of \( z_H(x_H, y_H) \) when \( \epsilon_A = -0.005 \) cm, highlighting a shift in the minimum point of \( z_H \) along the \( x \)-axis and anisotropic deformation around this point.

FIG.~\ref{fig:15tilt}(b) depicts the relationship between \( \epsilon_A \) and the \( x \)-coordinate of \( z_H \)'s minimum point \( x_{\text{min}} \). Lowering point \( A \) shifts \( x_{\text{min}} \) in the positive \( x \)-direction, while raising it shifts \( x_{\text{min}} \) in the negative \( x \)-direction.

FIG.~\ref{fig:15tilt}(c) demonstrates how \( \epsilon_A \) influences the second derivatives \( \partial^2 z_H/{\partial x_H}^2 \) and \( \partial^2 z_H/{\partial y_H}^2 \), making the behavior around the minimum point non-isotropic. Specifically, when \( \epsilon_A < -0.012 \) cm, the minimum point becomes a saddle point.

These results are consistent with the gliding plane shown in FIG.~\ref{fig:05z_contour} after introducing a tilt by $\epsilon_A$, that is, by adding an inclination to $z_H(x_H, y_H)$ in FIG.~\ref{fig:05z_contour}.

\subsubsection{Effects of Wire Length Discrepancies}
\begin{figure}
\includegraphics[width=1\columnwidth]{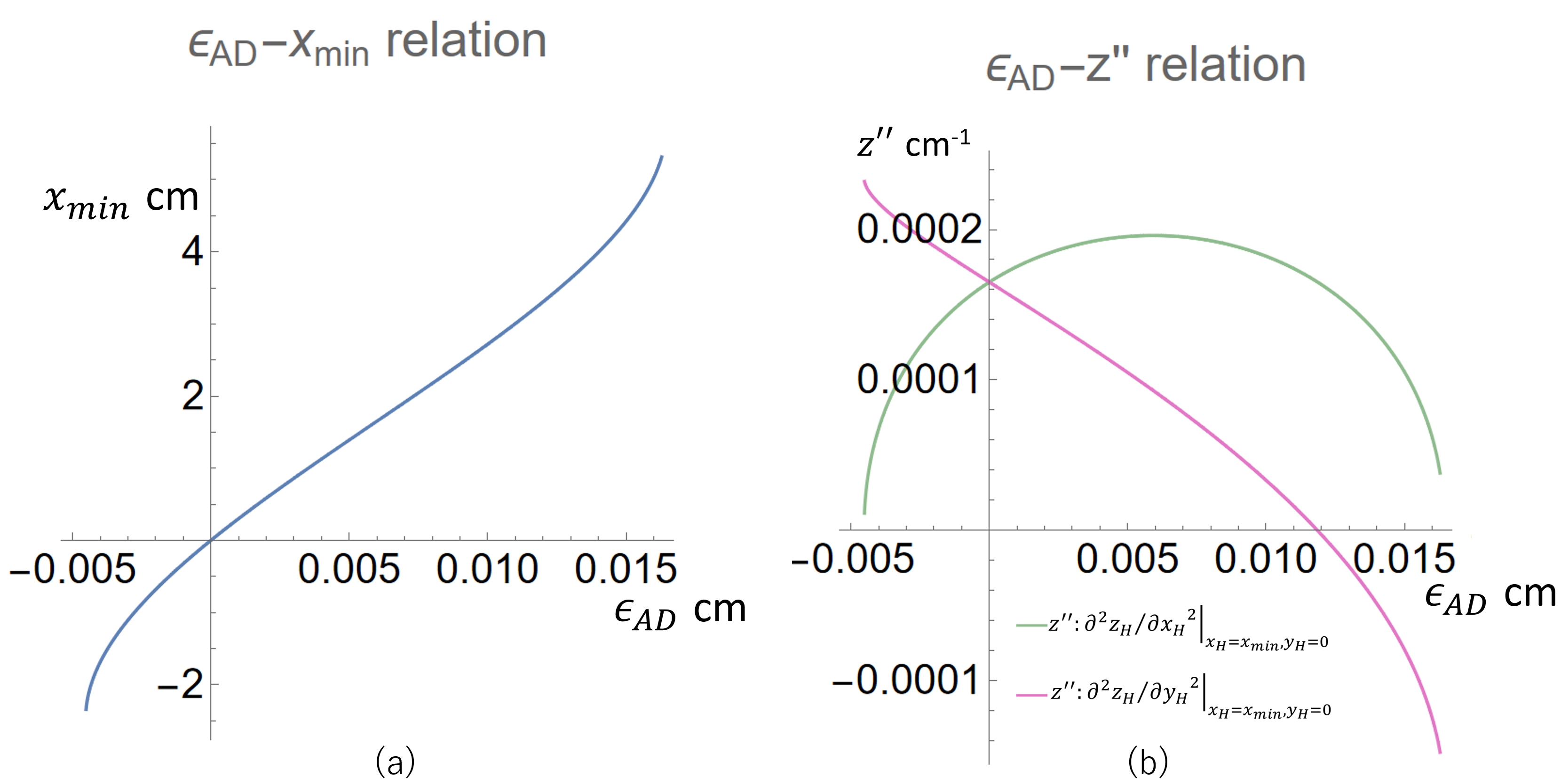}
\caption{\label{fig:16length} Effects of Wire Length Discrepancies: 
Panel (a) depicts the relationship between the wire length change \(\epsilon_{AD}\) and the x-coordinate of the \(z_H\) minimum point, \(x_{\text{min}}\). Increasing the length of wire AD shifts the minimum point in the +x direction, while decreasing it shifts the point in the -x direction. Panel (b) illustrates the impact of \(\epsilon_{AD}\) on the second derivatives \(\partial^2 z/{\partial x_H}^2\) and \(\partial^2 z/{\partial y_H}^2\) at the \(z_H\) minimum point, highlighting how changes in wire length affect the curvature of the surface at the minimum point.}
\end{figure}

This section examines the behavior of the PDRL when there is a slight deviation in one of its wire lengths. Normally, the length of wire AD should be \( l \equiv \sqrt{(a/2)^2 + b^2} \); however, we consider it to be \( l + \epsilon_{AD} \) and analyze the implications of this minor change. The results are presented in FIG.~\ref{fig:16length}.

 In FIG.~\ref{fig:16length}, panel (a) shows how a change in \( \epsilon_{AD} \) affects the position of the minimum \( z_H \), plotting \( \epsilon_{AD} \) against the x-coordinate \( x_{min} \) of the minimum point. Panel (b) demonstrates the relationship between \( \epsilon_{AD} \) and the second derivatives \( \partial^2 z_H/{\partial x_H}^2 \) and \( \partial^2 z_H/{\partial y_H}^2 \) at the minimum point. 

\subsubsection{Effects of the COM's Horizontal Displacement}

\begin{figure}
\includegraphics[width=1\columnwidth]{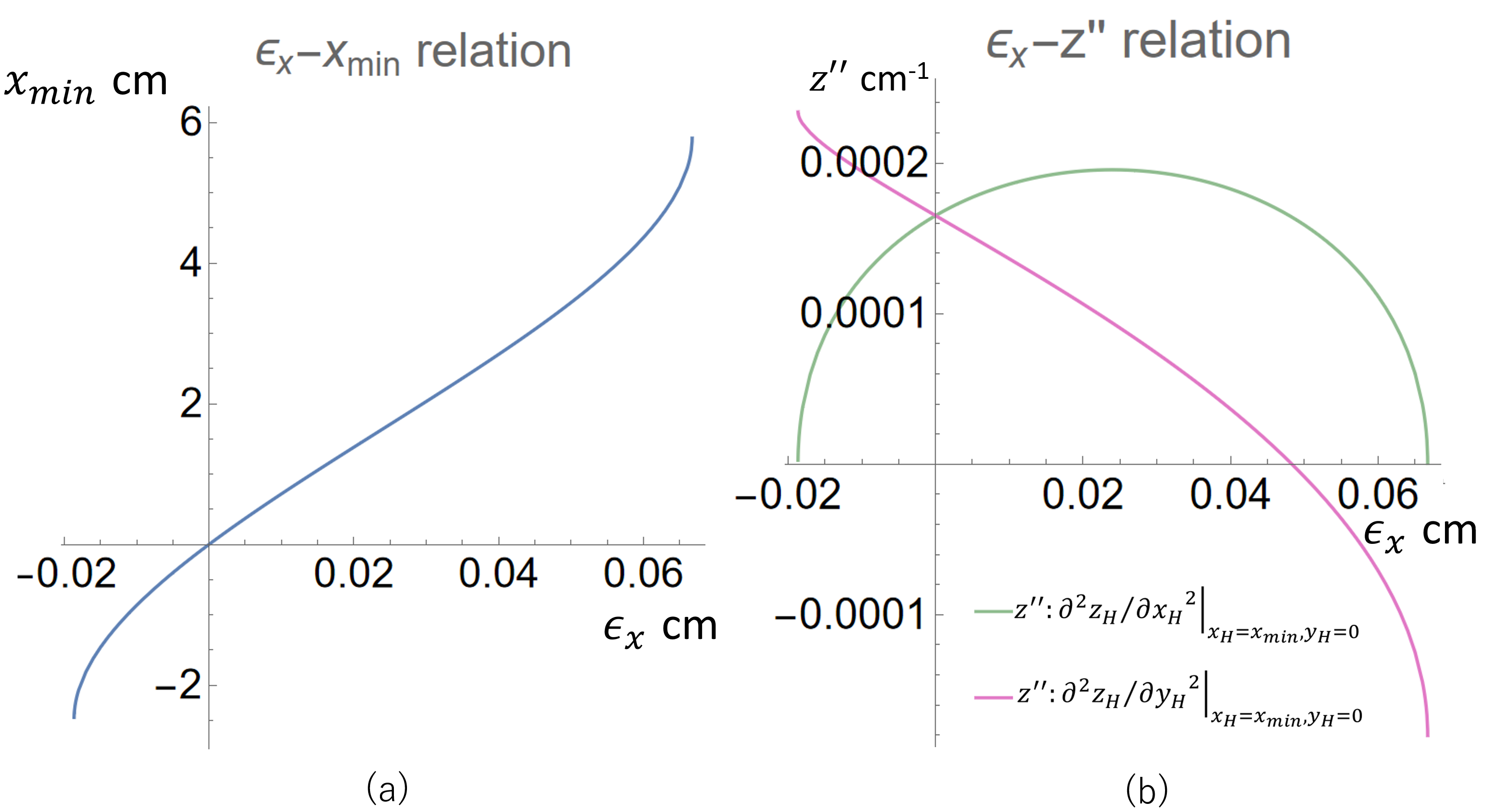}
\caption{\label{fig:17deaxis}
Effects of COM's Horizontal Displacement: 
Panel (a) illustrates the relationship between the horizontal displacement of the COM \(\epsilon_x\) and the x-coordinate of the minimum point of \(z_H\), denoted as \(x_{\text{min}}\). A shift of the COM in the +x direction results in a corresponding shift of the minimum point along the +x axis.
Panel (b) demonstrates the effects of \(\epsilon_x\) on the second derivatives \(\partial^2 z_H/{\partial x_H}^2\) and \(\partial^2 z_H/{\partial y_H}^2\) at the minimum point of \(z_H\). It highlights how displacements in the COM influence the anisotropic deformation characteristics around the minimum point.
}
\end{figure}

For the PDRL, FIG.~\ref{fig:17deaxis} shows behavior ,in the same manner as FIG.~\ref{fig:16length}, when the COM, normally positioned at \((0, 0, c)\) in a state where the base plane is horizontal, is displaced to \((\epsilon_x, 0, c)\).
Panel (a) illustrates the displacement of the minimum point of \(z_H\) in response to \(\epsilon_x\), plotting the relationship between \(\epsilon_x\) and the \(x\)-coordinate of \(z_H\)'s minimum point, \(x_{\text{min}}\). 
Panel(b) displays the relationship between \(\epsilon_x\) and the second derivatives \(\partial^2 z_H/{\partial x_H}^2\) and \(\partial^2 z_H/{\partial y_H}^2\) at the minimum point of \(z_H\). 

Even a slight shift of the COM in the positive x direction results in a substantial corresponding shift of \(x_{\text{min}}\) in the same direction.

\subsubsection{\label{sec:level3}Compensation through Wire Length Adjustments}
\begin{figure}
\includegraphics[width=1\columnwidth]{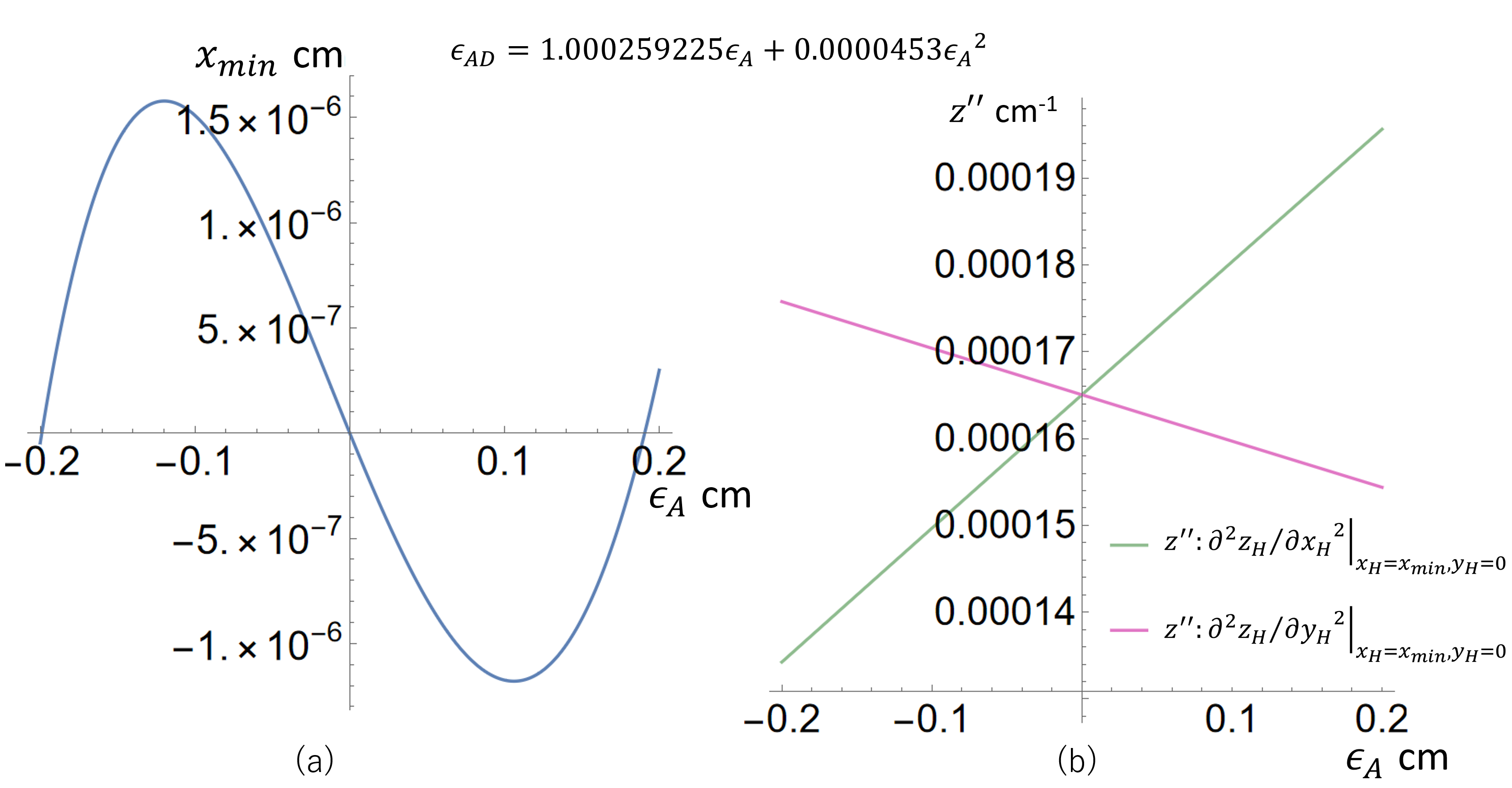}
\caption{\label{fig:18tilt-length}
Adjusting Wire Length to Compensate for Tilt Effects: 
This figure illustrates how the effects of tilt ($\epsilon_A$) are canceled by small adjustments to the wire length $\epsilon_{AD}$, given by $\epsilon_{AD}=1.000259225\epsilon_A+0.0000453{\epsilon_A}^2$.
Panel (a) shows that specific adjustments to \(\epsilon_{AD}\) can effectively maintain the position of the \(z_H\) minimum point within a range of \(\pm 2 \times 10^{-6}\) cm, despite variations in \(\epsilon_A\). Panel (b) highlights that, although the minimum position can be maintained centrally, the second derivatives \(\partial^2 z_H / {\partial x_H}^2\) and \(\partial^2 z_H / {\partial y_H}^2\) vary, indicating that the shape around the minimum point does not retain isotropy.}
\end{figure}

\begin{figure}
\includegraphics[width=1\columnwidth]{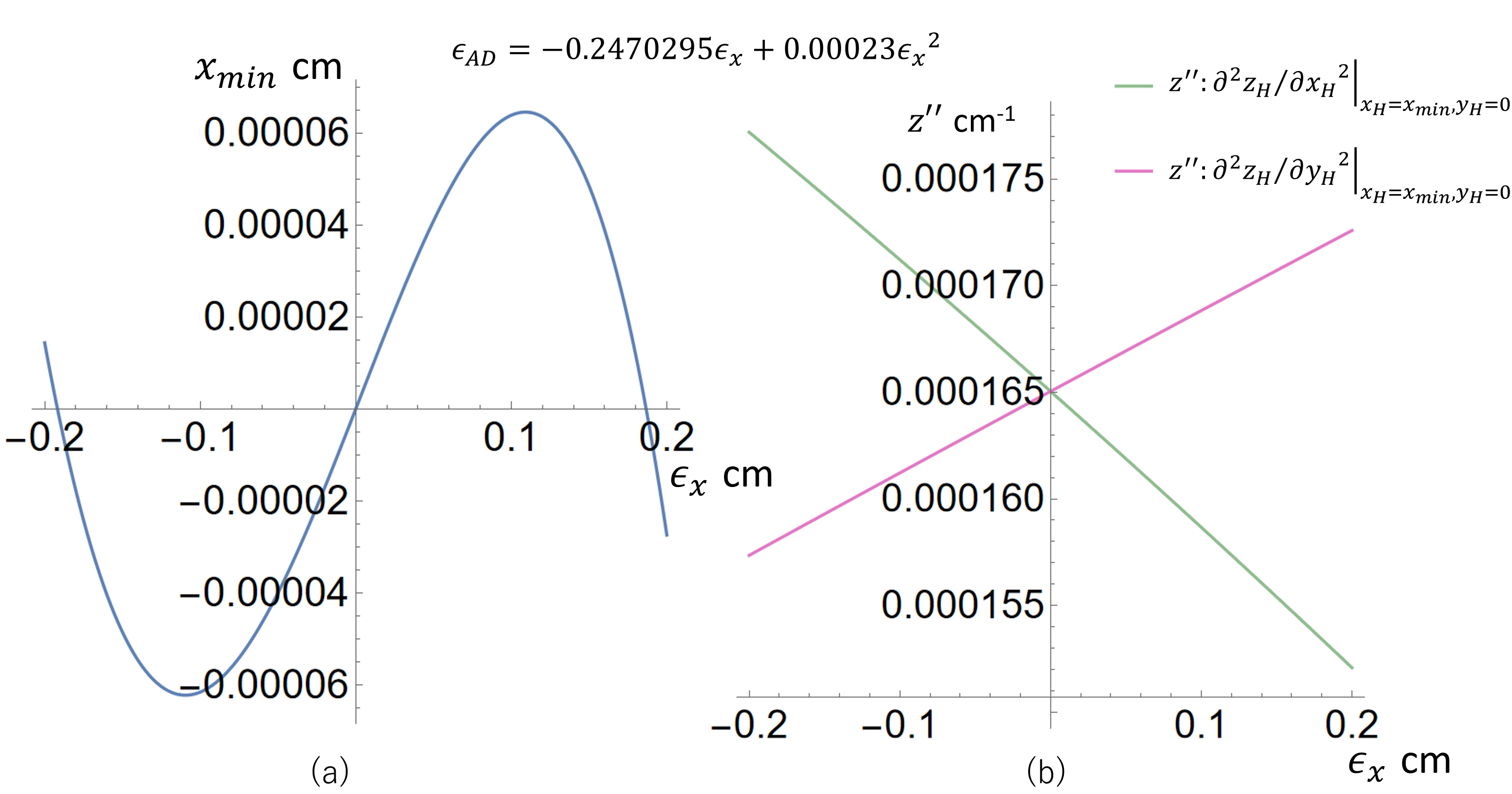}
\caption{\label{fig:19tilt-deaxis}
Adjusting Wire Length to Compensate for COM's Horizontal Displacement: 
This figure illustrates how the effects of COM's Horizontal Displacement ($\epsilon_x$) are canceled by small adjustments to the wire length $\epsilon_{AD}$, given by $\epsilon_{AD}=-0.2470295\epsilon_x+0.00023{\epsilon_x}^2$.
Panel (a) shows that appropriate adjustments to \(\epsilon_{AD}\) can effectively maintain the position of the \(z_H\) minimum point within a range of \(\pm 6 \times 10^{-5}\) cm, despite variations in \(\epsilon_x\). Panel (b) highlights that, although the minimum position can be maintained centrally, the second derivatives \(\partial^2 z_H / {\partial x_H}^2\) and \(\partial^2 z_H / {\partial y_H}^2\) vary independently, indicating that the shape around the minimum point does not retain isotropy.}
\end{figure}

As we have observed so far, inclination $\epsilon_A$, wire length discrepancies $\epsilon_{AD}$, and the COM’s horizontal displacement $\epsilon_x$ exhibit similar characteristics, which leads to the consideration of canceling them out mutually. Adjusting wire length through heating has been previously studied \cite{THR}, and here, we explore the possibility of compensating for the inclination $\epsilon_A$ and COM’s horizontal displacement $\epsilon_x$ by adjusting the wire length $\epsilon_{AD}$. Specifically, for $\epsilon_A$ and $\epsilon_x$ values of -0.2 cm, 0 cm, and +0.2 cm, we determine the $\epsilon_{AD}$ that makes the minimum point of $z_H$ satisfy $x_H = 0$, placing it on the z-axis. The results of compensating using a quadratic function interpolated between these three points for $\epsilon_{AD}$ are shown in FIG.\ref{fig:18tilt-length} and FIG.\ref{fig:19tilt-deaxis}. In each figure, the equation shown to determine $\epsilon_{AD}$ is the quadratic interpolation function used for compensation. Panel (a) in each figure displays $x_{\text{min}}$ after compensation using this interpolation, and in both cases, $x_{\text{min}}$ is found to be very small, indicating that the compensation is effective.
However, even in such cases, as shown in Panel (b), the values of the second derivatives \( \partial^2 z_H/{\partial x_H}^2 \) and \( \partial^2 z_H/{\partial y_H}^2 \) diverge, indicating that isotropy cannot be maintained at the minimum point.

\section{discussion}
The discussions so far have proceeded on the premise that the minimum possible value of \(z_H\) is considered for given horizontal positions \(x_H, y_H\) of the COM. However, the rigid body also has a degree of freedom in torsion \(\varPhi\), and its position is defined within this three-dimensional phase space.

Nevertheless, when used as a vibration isolation system (VIS) for gravitational wave detectors, the following considerations apply:
\begin{itemize}
    \item The system is typically operated very close to the most stable equilibrium point, $x_H=0,y_H=0,\varPhi=0$.
    \item The motion that requires isolation is translational, and no external forces are applied that would induce rotational motion in the $\varPhi$ direction.
    \item According to Equation (\ref{eq16}), minimal motions in the $x_H, y_H$ directions do not induce rotations in $\varPhi$, as $\varPhi = 0$ corresponds to the position of the lowest COM and the minimum potential energy.
\end{itemize}
Considering these factors, the premise that \(z_H\) attains its minimum value as described in this article is justified and sufficient for practical purposes.

The three-point RL exhibits isotropy at infinitesimal amplitudes, and its equivalent pendulum length \( r_0 \) is given by
\begin{eqnarray} r_0 = \frac{1}{d^2z_H/{dr_H}^2} = \frac{4b(2b+c)^2}{a^2-4bc} .\label{eq24}\end{eqnarray}
At \( c = 0 \), \( r_0 \) becomes \( 16b^3/a^2 \),which implies that it scales with dimension in similar shapes. However, when the lateral dimension $a$ is constant, \( r_0 \) is proportional to the cube of $b$, and it's easy to extend the equivalent pendulum length by adjusting $c$.

As \( \varPsi \rightarrow \theta \) when \( r \rightarrow 0 \), it can be considered that the direction in which the axis tilts during infinitesimal amplitude oscillations is the same as the direction of these oscillations. The rotational motion determined by this tilting is governed by the moment of inertia, which can be assessed from the COM oscillation by
\begin{eqnarray}
    \frac{d\varTheta}{dr_H} = \frac{1}{2b+c}.
\end{eqnarray}
There's no need to consider torsion around the rigid body axis.

This article presents a theoretical examination under the assumption that there is no deformation of the rigid body or extension and vibration of the wires. Therefore, in actual devices where such effects cannot be neglected due to the material properties, mass, and dimensions of the device, including the wire, it is necessary to consider appropriate corrections. In practice, it is assumed that the linkage is constructed by clamping both ends of the wire to the support and the rigid body, with the wire bending at the clamp points. The impact of wire extension becomes more significant as the wire diameter decreases. On the other hand, as the wire becomes thicker, the displacement of the pivot point increases, requiring careful attention.

\section{conclusion}
This study elucidated the equivalent pendulum length necessary for using the three-point RL as a VIS, as well as the rigid body rotation corresponding to oscillations. The influence of structural variations were also clarified.

In the future, key theoretical challenges will include addressing wire elongation and flexure correction. On the implementation side, the development of methods to correct and control the sensitive parameters identified in this study will be a critical focus.

\begin{acknowledgments}
This research was supported by International Graduate Program for Excellence in Earth-Space Science (IGPEES), a World-leading Innovative Graduate Study (WINGS) Program, the University of Tokyo.
\end{acknowledgments}

\nocite{*}
\bibliography{aipsamp}

\end{document}